# Demythization of XML Query Processing: Technical Report

Petr Lukáš #     Radim Bača #     Michal Krátký #     Tok Wang Ling *
# Department of Computer Science, VŠB – Technical University of Ostrava, Czech Republic
* Department of Computer Science, National University of Singapore
{petr.lukas,radim.baca,michal.kratky}@vsb.cz, lingtw@comp.nus.edu.sg

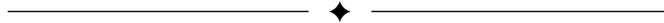

**Abstract**—XML queries can be modeled by twig pattern queries (TPQs) specifying predicates on XML nodes and XPath relationships satisfied between them. A lot of TPQ types have been proposed; this paper takes into account a TPQ model extended by a specification of output and non-output query nodes since it complies with the XQuery semantics and, in many cases, it leads to a more efficient query processing. In general, there are two types of approaches to process a TPQ: holistic joins and binary joins. Whereas the binary join approach builds a query plan as a tree of interconnected binary operators, the holistic join approach evaluates a whole query using one operator (i.e., using one complex algorithm). Surprisingly, a thorough analytical and experimental comparison is still missing despite an enormous research effort in this area. In this paper, we try to fill this gap; we analytically and experimentally show that the binary joins used in a fully-pipelined plan (i.e., the plan where each join operation does not wait for the complete result of the previous operation and no explicit sorting is used) can often outperform the holistic joins, especially for TPQs with a higher ratio of non-output query nodes. The main contributions of this paper can be summarized as follows: (i) we introduce several improvements of existing binary join approaches allowing to build a fully-pipelined plan for a TPQ considering non-output query nodes, (ii) we prove that for a certain class of TPQs such a plan has the linear time complexity with respect to the size of the input and output as well as the linear space complexity with respect to the XML document depth (i.e., the same complexity as the holistic join approaches), (iii) we show that our improved binary join approach outperforms the holistic join approaches in many situations, and (iv) we propose a simple combined approach that utilizes advantages of both types of approaches.

## 1 INTRODUCTION

Searching for occurrences of a twig pattern in an XML tree is often considered as a core problem of the structural XML query processing. An enormous research effort has been devoted to this problem [3] and we can identify two major structural XML query processing approaches: (1) binary join approaches [1], [36], [2], [30], and (2) holistic join approaches [7], [8], [9], [18], [20], [14], [17], [19], [4]. Whereas the binary join approach builds a query plan as a tree of interconnected binary operators, the holistic join approach evaluates a whole query using one operator (i.e., using one complex algorithm). Both types of approaches work with the same access path (i.e., an inverted list where the XML nodes are sorted according to the document order) and they have its pros and cons which are described more thoroughly in Section 4. After two decades of research in structural XML query processing, can we decide which approach to use and when? Actually, the only comparison between both types of approaches can be found in [7]; however, it was an early stage of the TPQ processing and several important advances of binary joins have occurred since then [2], [36], [13], [16], [23], [30]. One of the most important advance in the field of binary joins is a so-called *fully-pipelined* (FP) query processing plan [36] since a join operation of such a plan does not wait for the complete result of the previous join operation, i.e., the join operation result is not materialized and no explicit sorting is used. Another important advance of the binary join approach is a utilization of semi-joins [2] whenever it is possible since they are computationally less demanding. There are more than thirty holistic joins in the literature where none of them is compared with a binary join. Vast majority use the early *TwigStack* work [7] as an argument that holistic joins are superior over binary joins and some of them ignore binary joins completely. As we show in our experiments, the binary join approach can outperform the holistic join approach for TPQs with a low selectivity if we use the advances in binary join query processing [36] which is in contrast to the result of [7]. Moreover, the advantages of binary join approaches become more evident if the non-output query nodes are considered.

One aim of this article is to compare both types of the approaches and to dispel the most common myths about binary joins claiming that (1) binary joins produce large intermediate results [7], [8], [9], [18], [20], [27], [19], [4], and (2) the space, I/O, and time complexity of the TPQ processing can be proved only for holistic joins. Especially the first myth is repeated in a major number of holistic join articles as the main argument to avoid a comparison with binary joins and in this article we show that it is not correct. We may use an analogy from the relational database world where everyone understands that a query plan using binary joins such as a nested-loop join is processed in a pipelined manner [12] and a new SQL processing algorithm can hardly avoid a comparison with them using the possibly large intermediate result as an argument.

The problem of searching for occurrences of a TPQ in an XML tree is very general and does not correspond to

2the XQuery language. The major drawback of the TPQ model is that it misses the semantics related to the 'for', 'let', and 'return' clauses, therefore, generalized twig pattern (GTP) [10] has been introduced. [23] and [4] have shown that mainly the semantics related to the 'for' clauses can be important for the efficiency of the query processing since it basically defines the number of result tuples. The ideas applied to consider the 'for' clause semantics can be easily generalized to the 'let' and 'return' clauses. In other words, we consider the GTP model too complicated for the comparison and we just slightly extend the TPQ model. We simply define two types of query nodes in a TPQ: (i) *output query nodes* corresponding to the 'for' clauses and (ii) *non-output query nodes* corresponding to all other query nodes. Let us note that the output nodes are sometimes called extraction points [23] or return nodes [8].

Let us note that the lexicographic order of the result is required by the XQuery data model, and therefore, a building of an FP plan for the TPQ that considers the non-output query nodes is not a trivial problem. In this article, we combine ideas proposed in [36], [2], [30] and add certain modifications of binary join approaches; in this way we introduce such a building of an FP plan for the TPQ. Moreover, we show that it is not necessary to utilize a cost-based optimizer to build an FP plan. Let us note that the holistic join approaches also do not often utilize a cost-based optimizer. As a result, we simply consider a cost-based optimization to be orthogonal to both types of approaches [15], [5]. In this way, we can do a fair comparison of both approaches since there is the same access path and no cost-based optimizer is used. As a bonus, we can prove that for a certain class of TPQs our binary join approach has the linear time complexity with respect to the size of the input and output and the linear space complexity with respect to the depth of the XML document. To the best of our knowledge, this is the first work which proves such a complexity of the binary join approaches.

The main contributions of this paper are summarized as follows:

1) We introduce our binary join approach based on FP plans which is an extension of the existing ones. It builds FP plans for a TPQ considering non-output query nodes without a cost-based optimizer. This contribution debunks the first myth related to the intermediate result of binary joins.
2) We prove time and space complexities of our binary join approach for a certain class of TPQs. We show that the complexity of the binary join approach bears many similarities to the complexity of holistic join approaches. The only difference is that the complexity of holistic approaches is not influenced by the number of non-output query nodes. This contribution debunks the second myth related to the complexity of binary joins.
3) We conduct a thorough experimental comparison of both approaches. We clearly describe use-cases in which each of them excels. We show that our binary join approach is more likely to be advantageous over the holistic join approach for TPQs with a higher ratio of non-output query nodes or with a higher query selectivity, and this experimental observation is in the compliance with the complexity analysis.
4) We propose a combined approach that utilizes the results of the comparison and allows us to use advantages of both types of approaches.

The structure of the paper is as follows. Section 2 depicts the TPQ processing problem, Section 3 provides several motivation examples and Section 4 gives an overview of TPQ processing techniques. Section 5 includes a thorough description of the binary join algorithms and Section 6 describes how to build an FP plan based on these algorithms. Section 7 conducts a thorough complexity analysis of our binary join approach. Section 8 experimentally compares the both types of approaches to distinguish their weak and strong aspects.

## 2 PRELIMINARIES

Let us first define some basic terms we use in this paper.

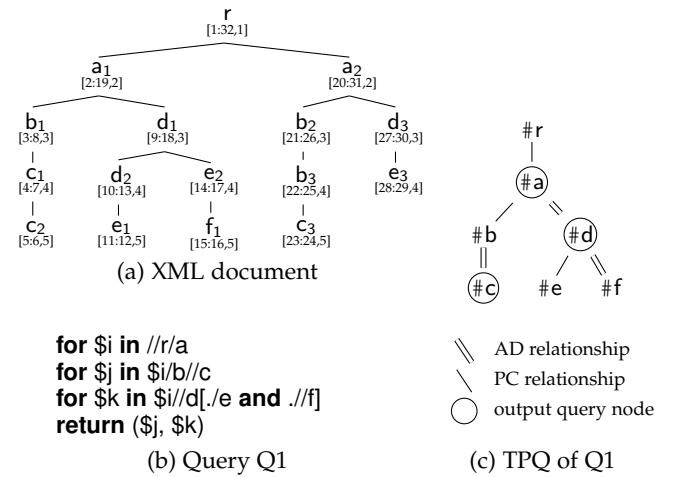

Fig. 1. A sample document, an XQuery query, and a TPQ

### 2.1 Document and Query Model

An *XML document* (or simply a *document*) is an ordered rooted tree. The nodes in the tree are called *data nodes*. Each data node is labeled with a *tag*. A sample XML document is depicted in Figure 1a. We depict the pre-order numbering of data nodes with the same tag in the subscripts (i.e., the subscripts are not a part of the tags). The tuples under the data nodes represent containment labels which are explained further in Section 2.4.

A *twig pattern query (TPQ)* is also an ordered rooted tree. The nodes in the tree are called *query nodes*. Each query node is labeled with: (1) an expression #$\theta$ specifying all data nodes with the tag $\theta$ and (2) information whether the query node is an *output query node* or a *non-output query node*, i.e., whether the query node corresponds to a 'for' clause or not, respectively. The edges represent XPath relationships; we consider only *ancestor-descendant* (AD) and *parent-child* (PC) relationships in this paper, as far as it is common for a lot of XML query processing approaches [1], [7], [30]. Single and double lines represent PC and AD edges, respectively, the output query nodes are circled. Since a TPQ models an XML query, we use the terms 'query' and 'TPQ' interchangeably. An example of a TPQ is illustrated in Figure 1c, the TPQ corresponds to the XQuery query Q1 in Figure 1b.

As we mentioned earlier, there are other query models covering the semantics of the TPQ with non-output query nodes in the literature such as GTP [10] or TP+Output [35] model. For the sake of simplicity, we do not utilize these richer models since our focus is only on the semantics related to the 'for' clause due to its importance for the query processing efficiency. Therefore, in this paper, the output query nodes are related only to the 'for' clauses. For example, the query Q1 has three output query nodes corresponding to the three 'for' clauses (even though there are only two variables in the 'return' clause). In Appendix D, we show that the GTP model, covering also the 'let'/'return' clauses and 'or'/'not' logical connectives, can be easily integrated into our binary join approach and it is actually implemented in our prototype [11].

## 2.2 Problem Statement

Let us have a document $D$ and a query $Q$ with $n$ query nodes $q_1, \ldots, q_n$. A *complete match* is an $n$-dimensional tuple $[d_1, \ldots, d_n]$, where $d_1, \ldots, d_n$ are data nodes from $D$ specified by the query nodes $q_1, \ldots, q_n$ such that the relationships between $d_1, \ldots, d_n$ satisfy the AD or PC relationships defined by edges between $q_1, \ldots, q_n$. An *output match* is a projection of a complete match such that data nodes corresponding to non-output query nodes are omitted. The result of the query $Q$ for the document $D$ is an ordered set of all output matches; these output matches are lexicographically sorted by the document order of data nodes corresponding to the output query nodes in a pre-order traversal.

The lexicographic sorting complies to the XQuery language specification [28] for most of XQuery queries processed in the *ordered* mode. However, there exist XQuery queries that cannot be modeled by a TPQ such that the result is ordered correctly. We assume that these queries are rather impractical and rare and we do not consider them in this paper. An example of such a query is shown in Appendix A.

## 2.3 Query Core

In this section, we define some specific parts of a query. The specific parts can be processed by specific algorithms as we shall see in Section 6. A *query core node* is an output query node or any query node lying on a path between two output query nodes. With respect to a query $Q$, a *query core* is a minimal subtree of $Q$ that encompasses all of the query core nodes of $Q$, each query has a unique core. A *constraining subquery* $\Pi_q$ for a query core node $q$ (called the *constrained node*) is a query subtree formed of $q$ and all non-output query nodes connected with $q$ by any edges (PC or AD) not lying on a path between two output query nodes. Putting the constraining subqueries and the query core together results in the original query. In Section 7, we show that the use of different join algorithms for the query core and for the constraining subqueries leads to a more efficient query processing.

*Example 2.1.* Let us consider the query Q1 in Figure 1c.
There are 4 query core nodes: #a, #b, #c, and #d (see Figure 2a). The constraining subqueries $\Pi_{\#a}$ and $\Pi_{\#d}$ are depicted in Figure 2b and 2c, respectively. The constraining subqueries $\Pi_{\#b}$ and $\Pi_{\#c}$ include only the query core nodes themselves.

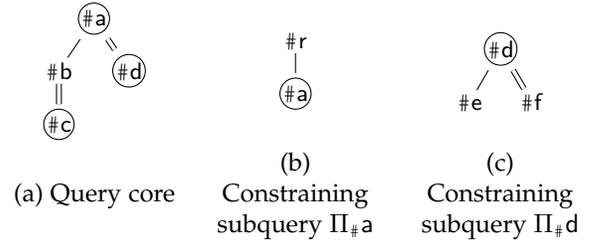

(a) Query core   (b) Constraining subquery $\Pi_\#$a   (c) Constraining subquery $\Pi_\#$d

Fig. 2. Subtrees of the query Q1

## 2.4 Labeling Scheme

Each data node in a document is labeled with a unique *node label* [1], [7], [4]. The node label allows us to resolve XPath relationships between data nodes. In this paper, we use the containment labeling scheme [38], where a node label is a 3-tuple $[left : right, level]$. Using the containment scheme, a data node $a$ is an ancestor of a data node $d$ if the interval $(a.left, a.right)$ contains the interval $(d.left, d.right)$. Similarly, a data node $p$ is the parent of a data node $c$ if $p$ is an ancestor of $c$ and $p.level + 1 = c.level$. Let us note that any labeling scheme which is capable of resolving AD or PC relationships (e.g., Dewey order [31]) can be also used. In Figure 1a, we see an XML tree with node labels of the containment labeling scheme. Further in the text, we use the terms 'data node' and 'node label' interchangeably.

## 2.5 Query Plan

An *n-ary operator* is a mapping $T_{in_1} \times \cdots \times T_{in_n} \to T_{out}$, where $T_{in_1}, \ldots, T_{in_n}$ are input streams and $T_{out}$ is the output stream. A *stream* $T$ is an abstract data structure supporting the following operations:

- `advance`$(T)$ – moves to the next item in $T$,
- `finished`$(T)$ – returns true, iff the whole stream $T$ has been completely iterated,
- `current`$(T)$ – returns the current item of $T$.

In addition, $|T|$ denotes the size of $T$, i.e., the number of possible `advance` invocations before $T$ is finished.

In this paper, we combine operators working with a *table* [1], [2], [4] and operators working with a *list* [30]; both data structures implement the stream. A table is an array of tuples of node labels lexicographically sorted by one or more columns using the document ordering (i.e., the $left$ value of the node labels), a list is an equivalent of a one-column table.

A *query plan* is an ordered rooted tree of operators where inputs of an operator are outputs of its children. A *fully-pipelined (FP) plan* [36] is a query plan where each operator $o$ produces the output stream sorted such that the parent of $o$ (or an end-user if $o$ is the root) requires. The FP plans bring the following advantages: (1) an explicit sorting is not necessary, and (2) the parent operator can consecutively read the result without waiting for the complete result of the previous operator (i.e., the operators can be implemented as cursors as it is common in relational DBMSs).

## 2.6 Query Processing Operators

In this section, we depict several query processing operators that we consider in our query plans.

### 2.6.1 Binary Join

In the literature, the binary join operator is sometimes called a structural [1], [36] or a containment join [2]. There are many types of this operator which can be categorized from various points of view. Following the specific parts of a query defined in Section 2.3, our approach uses *partial-joins* [2] to process the query core and *semi-joins* [30] to process the constraining subqueries. The major difference of semi-joins is that they work only with lists instead of tables. Their implementation is, therefore, more efficient, but they cannot be used to process the query core.

The *partial-join* operator [2] is defined as follows:

$$T_o = PartialJoin_{[M_a, M_d, i, j, \alpha]}(T_a, T_d),$$

where $T_a$ and $T_d$ are tables with $m$ and $n$ columns, $M_a$ and $M_d$ are binary *projection masks* with $m$ and $n$ bits specifying which columns from $T_a$ and $T_d$ are projected to the output, $i$ and $j$ are indexes of *join columns* in $T_a$ and $T_d$, respectively. $\alpha$ is an AD or PC relationship. The partial-join operator requires $T_a$ and $T_d$ to be sorted by the join columns. $T_o$ is the output table with $o$ columns ($o = m' + n'$), where $m'$ and $n'$ are the numbers of non-zero bits in $M_a$ and $M_d$, respectively. $T_o$ is computed by joining all tuples in $T_a$ and $T_d$ by the join columns satisfying the relationship $\alpha$. There are two variants of partial-join: *ancestor-sorted* and *descendant-sorted*; they output $T_o$ lexicographically sorted by data nodes from the join columns of $T_a$ or $T_d$, respectively.

The *semi-join* operator [30] is defined as:

$$T_o = SemiJoin_{[\alpha]}(T_a, T_d),$$

where $T_a$ and $T_d$ are lists and $\alpha$ is an AD or PC relationship. There are two variants of the semi-join: (i) an *ancestor-filtering* variant produces a list $T_o$ of all data nodes $a \in T_a$ such that there is a data node $d \in T_d$ in the relationship $\alpha$ with $a$, and (ii) a *descendant-filtering* variant produces a list $T_o$ of all data nodes $d \in T_d$ such that there is a data node $a \in T_a$ in the relationship $\alpha$ with $d$.

### 2.6.2 Holistic Join

A holistic join is an operator with $n$ input lists $T_{q_1}, \ldots, T_{q_n}$ of data nodes producing a table of all output matches [7]. When compared to a binary join, the major difference is that a holistic join considers current items of all streams in every step of the algorithm, whereas a binary join considers only a pair of streams.

### 2.6.3 Index Scan

Both the binary and the holistic joins work with streams of data nodes retrieved by an operator called *IndexScan* (IS). $IS(q)$ retrieves data nodes specified by $q$, where $q$ is a tag. We consider an index data structure implementing this operator (e.g., B$^+$–tree based inverted list). Let us note that the usage of such a data structure is common in XML DBMSs [1], [7], [4], [27], [30].

## 2.7 Query Processing Approaches

Following the operators described in the previous subsection, we can distinguish three TPQ processing approaches: (i) *binary join approaches*, (ii) *holistic join approaches* and (iii) *combined join approaches*. The binary join approaches utilize a query plan with a number of *binary join* operators where each join is related to one edge in the query. The holistic join approaches utilize a query plan with one operator – a *holistic join*. In combined join approaches, both binary joins and holistic join operators are combined in a single query plan.

## 3 MOTIVATION EXAMPLES

In this section, we provide two complex motivation examples to intuitively demonstrate our binary join approach based on FP plans and to compare binary and holistic join approaches.

### 3.1 Intuitive Example of our Binary Join Approach

The most of techniques used in our approach have been already proposed in the literature [1], [36], [2], [30], but they have not been used together to process any TPQ with defined non-output query nodes using FP plans such that the result is lexicographically sorted as it is required by XQuery. Let us recall, that in the FP plan, no explicit sorting is necessary and, therefore, any intermediate result does not have to be materialized. We describe the algorithm to build an FP plan thoroughly in Section 6 in more detail.

Considering the sample XML document in Figure 1a and the TPQ in Figure 1c, our approach processing the TPQ is logically done in two phases: (1) filtering and (2) joining. In the filtering phase, we process the constraining subqueries such that we first filter out all data nodes of #a whose parent does not match #r. Once we have the filtered data nodes of #a, their parents (data nodes of #r) are no longer useful and can be omitted, i.e., we can use a binary semi-join. Similarly, we restrict the data nodes of #d such that we first filter out data nodes of #d not having a child of #e and, subsequently, we filter out the data nodes of #d not having a descendant of #f. Obviously, these two filterings can be also done in the reversed order. For the rest of the query processing, the data nodes of #e and #f can be omitted, i.e., semi-joins can be used again. The first phase produces restricted sets of data nodes for #a: ($a_1$, $a_2$) and for #d: ($d_1$).

In the joining phase, we first join data nodes of #b and #c to make pairs $[b, c]$, where $b$ is an ancestor of $c$. We output the pairs such that they are lexicographically sorted by $(c, b)$ automatically (i.e., we use a descendant-sorted partial join), we get the following pairs: $[b_1, c_1]$, $[b_1, c_2]$, $[b_2, c_3]$, $[b_3, c_3]$. After that, we join the restricted data nodes of #a with the pairs $[b, c]$ to make triplets $[a, b, c]$ such that $a$ is an ancestor of $c$ and the triplets are lexicographically sorted by $(a, c, b)$ (i.e., we use an ancestor-sorted partial join), the result includes the triples: $[a_1, b_1, c_1]$, $[a_1, b_1, c_2]$, $[a_2, b_2, c_3]$, $[a_2, b_3, c_3]$. Next, we have to test, whether in the triplets, $b$ is a child of $a$; all triplets that do not satisfy this property (i.e., $[a_2, b_3, c_3]$) are removed. We call it the *secondary relationship test* (see Section 5.1). At this point, we can omit the $b$ data nodes, since they are no longer useful. Since the triplets are sorted by $(a, c, b)$, removing the b data nodes would produce pairs sorted by $(a, c)$. Now, we can join these pairs with the restricted set of data nodes of #d using an ancestor-sorted partial join to make triplets $[a, c, d]$ lexicographically sorted by that order: $[a_1, c_1, d_1]$ and $[a_1, c_2, d_1]$. Let us note that the triplets are lexicographically sorted by output query nodes in a pre-order traversal corresponding to the order of 'for' clauses in the query. The part of a query processed in the second phase corresponds to the query core. Since the $i variable is not included in the 'return' clauses, we simply omit the $a_1$ data node to get the final result: $[c_1, d_1]$ and $[c_2, d_1]$.

## 3.2 Comparison of Binary and Holistic Join Approaches

Let us now provide a motivation example to demonstrate that the both binary and holistic join approaches have their pros and cons, which means that the holistic joins are not always superior to the binary joins. Let us consider an XML document with a simple structure in Figure 3a and two sample queries Q2 and Q3 in Figures 3b and 3c, respectively. Q2 returns $n$ triplets $[a_1, b_1, c_1], \ldots, [a_n, b_n, c_n]$ and Q2 returns one 4-tuple $[a_n, b_n, c_n, d]$. While Q2 is extremely low-selective since it returns almost all data nodes of the document, Q3 is extremely high-selective since it returns only the data nodes of the $a_n$'s subtree.

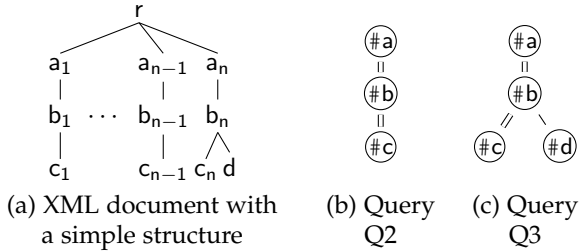

(a) XML document with a simple structure  (b) Query Q2  (c) Query Q3

Fig. 3. XML document and queries for the motivation example

In a nutshell, a typical holistic join approach processes a query as a whole using a chain of linked stacks constructed by a recursive function (e.g., `getNext()` in [7]). The recursive function filters out data nodes not participating any complete match. On the other hand, each binary join approach processes a query using a query plan composed of binary join operators. A typical binary join algorithm uses a simple loop to iterate through the both of its inputs. Similarly as a holistic join, it usually also utilizes a stack (see Section 5). Since in this paper we consider only the FP query plans (see Section 2.5), the intermediate results of individual binary join operators do not need to be materialized (i.e., stored in an intermediate data structure) or sorted (i.e., the output of each operator is already sorted such that the next operator requires).

Although the principle of both approaches is different, let us discuss several statistics of processing Q2 and Q3 using a binary and a holistic join approach. The statistics are as follows: (i) the number of invocations of the recursive function of a holistic join approach ($getNext_{HJ}$), (ii) the number of iterations in the loops of a binary join approach ($next_{BJ}$), and (iii) the number of stack operations (push / pop) for both approaches ($sOp_{HJ}$ and $sOp_{BJ}$). Our experiments show that these statistics represent the most time-expensive operations during the query processing. Let us analyze several cases of processing the both sample queries:

1) All query nodes are output and we use the *TwigStack* [7] holistic join and the *StackTreeAnc* [1] binary partial-joins.
2) All query nodes are output and we use the *GTPStack* [4] holistic join and the *StackTreeAnc* [1] binary partial-joins.
3) Only #a is an output query node (all other query nodes are non-output) and we use the *GTPStack* [4] holistic join and binary semi-joins [2], [30].

The case 1 represents an early stage of the XML query processing, i.e., the first proposed holistic join *TwigStack* and the traditional binary partial-join *StackTreeAnc* are used. In the case 2, we use a state-of-the-art holistic join *GTPStack* instead of *TwigStack* and in the case 3, we use a binary semi-join, since non-output query nodes are considered.

Let us first focus on the low-selective query Q2; the statistics are given in Table 1. In the case 1, $getNext_{HJ}$ is $4.5\times$ higher than $next_{BJ}$ which means that the recursive function of *TwigStack* is called unnecessarily too many times since it cannot filter out any data nodes. $sOp_{HJ}$ is approximately $1.5\times$ higher compared to $sOp_{BJ}$ since the binary join approach does not need to push/pop the leaf data nodes of #c to any stack. In the case 2, we use a state-of-the-art holistic join *GTPStack*, where $getNext_{HJ}$ is reduced, but it is still higher than $next_{BJ}$. $sOp_{BJ}$ and $sOp_{HJ}$ remain the same as in the case 1 since all query nodes are output. In the case 3, we employ the advantages of the binary semi-joins, since only one of the query nodes is output. We can see that $sOp_{BJ}$ is reduced to zero.

For the high-selective query Q3, we can observe that the state-of-the art holistic join algorithm (cases 2 and 3) *GTPStack* reduces many unnecessary invocations of the recursive function; $getNext_{HJ} \ll next_{BJ}$. Since the recursive function filters out useless data nodes (data nodes not participating the result), $sOp_{HJ}$ is also reduced compared to $sOp_{BJ}$ in the cases 1 and 2. In the case 3, BJ still does not utilize any stacks, which means that $sOp_{BJ} = 0$.

| query | case | $getNext_{HJ}$ | $next_{BJ}$ | $sOp_{HJ}$ | $sOp_{BJ}$ |
|---|---|---|---|---|---|
| Q2 | 1 | $9n$ | $2n+3$ | $6n$ | $4n$ |
|  | 2 | $5n+2$ | $2n+3$ | $6n$ | $4n$ |
|  | 3 | $3n+2$ | $2n+1$ | $2n$ | $0$ |
| Q3 | 1 | $4n+12$ | $n+6$ | $6$ | $2n+4$ |
|  | 2 | $9$ | $n+6$ | $8$ | $2n+4$ |
|  | 3 | $5$ | $n+4$ | $2$ | $0$ |

TABLE 1
Statistics of the comparison of holistic and binary join approaches

The example shows that binary join approaches can be advantageous compared to holistic join approaches for low-selective queries and for queries with a high ratio of non-output query nodes. We use these observations to propose a binary join approach that overcomes state-of-the-art holistic joins in these situations.

## 4 RELATED WORK

Various TPQ processing approaches have been developed during the last two decades [1], [7], [8], [9], [18], [20],

[38], [17], [4]. Following the previous section, two major types of XML processing approaches are represented by the binary join approaches [1], [38], [37], [36], [2] and the holistic join approaches [7], [8], [9], [18], [20], [17], [4]. The binary join approaches can be easily integrated into a full-fledged XQuery processor in order to support all XPath axes [10], [25], [6]. In [1], the *StackTreeAnc* and *StackTreeDesc* algorithms are introduced; these algorithms are based on several data structures: a stack and so-called self- and inherited- lists and they are considered as traditional binary joins used by many approaches [2], [30], [36]. In Section 5.1, we present an extension of *StackTreeAnc* (we call it *StackTreeAncSrt*) allowing us to build an FP plan for a TPQ considering the non-output query nodes semantics. In [2], the authors introduce a partial-join which can project out some columns of input tables. Their experiments show that the incorporating of the partial-join can reduce the query processing time. In [30], a structural semi-join operator is introduced; the semi-join is a special case of the partial-join. They present algorithms *NStack-SSJoin-Desc* and *NList-SSJoin-Anc* that implement the descendant-filtering and ancestor-filtering semi-joins, respectively. These algorithms are computationally less expensive than the partial-joins *StackTreeAnc* and *StackTreeDesc*, since *NStack-SSJoin-Desc* is implemented without any stack and *NList-SSJoin-Anc* is implemented without self- and inherited-lists (but the stack is still necessary). In Section 5.2, we introduce a stack-less variant of the ancestor-filtering semi-join.

In [36], the authors put forward several cost-based algorithms to select a query plan that is close to the optimal query plan. Their experiments show that the FP plan provides the most efficient performance when the complete query processing time is considered. The authors also show how to build the FP plan producing a table sorted by an arbitrary query node; however, they do not consider non-output query nodes and lexicographic sorting by multiple columns which is necessary to support the XQuery semantics. In Section 6, we show how to build the FP plans that overcome these limitations. To the best of our knowledge, this is the first approach where the lexicographic sorting is considered without an explicit sorting of the result.

Holistic join approaches provide precise theoretical guarantees [7] for a certain types of TPQ, they can be integrated into an XQuery algebra [24], and they can be even combined with binary join approaches in the same query plan [33], [34]. However, the number of XQuery constructs that can be processed holistically is limited, e.g., only some XPath axes are supported. There are many holistic join algorithms that can be categorized from different perspectives. The first perspective relates to the way the stacks and an intermediate result data structure are used: (1) the top-down holistic joins [7], [18], [20] usually skip most of irrelevant nodes before they are stored on stacks, (2) the bottom-up holistic joins [8] usually skip most of irrelevant nodes when they are popped out from their stacks, and (3) the combined holistic joins [17], [19], [4] employ advantages from both types of holistic joins. The second perspective categorizes holistic joins according to a query model used: (1) approaches that do not reflect the non-output query nodes [7], [18], [20], [17], and (2) approaches that do [8], [4].

We pick the *GTPStack* holistic join [4] as a representative of the state-of-the-art holistic algorithms [17], [19], [4] in our experimental comparison due to the fact that it combines the advantages of top-down and bottom-up approaches and it reflects the non-output query node semantics.

A combined join approach (i.e., an approach combining holistic and binary joins) is presented in [33], however it is not clear whether the query plans combining holistic and binary joins overcome query plans when only a holistic join is used. In Section 8.5, we propose a simple combined approach that outperforms both binary and holistic approaches under certain conditions.

## 5 BINARY JOIN ALGORITHMS

In this section, we describe algorithms implementing the partial-join and semi-join operators with focus on several improvements enabling us to build an FP plan when non-output query nodes are considered. A thorough description how to build an FP plan is given in Section 6.

The partial-join algorithms *StackTreeAnc* and *StackTreeDesc* were introduced in [1] and a few modifications regarding their signature were presented in [2]. The semi-join algorithms borrow ideas from [30], [2]. In the case of the partial-joins, the (AD or PC) XPath relationship is just a parameter of the algorithms[1]. For the semi-joins, both XPath relationships are implemented using different algorithms, since the processing of the AD relationship can be done more efficiently than the PC relationship. The partial- and semi- join algorithms used in our approach are summarized in Table 2.

| Join type | Relationship | anc.-sorted anc.-filtering | desc.-sorted desc.-filtering |
|---|---|---|---|
| part.-join | AD, PC | *StackTreeAnc* [1] | *StackTreeDesc* [1] |
| semi-join | AD | *SemiJoinAncAD* | *SemiJoinDescAD* |
|  | PC | *SemiJoinAncPC* | *SemiJoinDescPC* |

TABLE 2
Summary of binary join algorithms

### 5.1 Partial-join Algorithms

The partial-join algorithms *StackTreeAnc* and *StackTreeDesc* are based on merging two streams of node labels to satisfy the AD or PC relationship. Both algorithms require a stack to keep a track of all ancestors of any data node from $T_a$ at any time of the query processing. *StackTreeAnc* also requires extra data structures to produce a sorted result. These data structures are called the self- and inherited- lists and each entry in the stack of *StackTreeAnc* is associated with one self- and one inherited-list. A detailed description of both algorithms is given in [1].

We can identify a sorting problem in the case of queries with non-output query core nodes. In Figures 4a and 4b, we see two similar queries Q4 and Q5. The only difference of Q5 is that #d is a non-output query node. Before we describe the sorting problem which appears in the case of Q5, let us show how the partial-joins are used in an FP plan to process Q4.

*Example 5.1.* The result of Q4 for the document in Figure 1a includes all triplets $[a, d, e]$ such that $a$ is an ancestor

---

1. Let us note that the suffixes 'Anc' and 'Desc' in the algorithm names indicate the output sorting, not the relationship they solve; i.e., both algorithms can solve the both AD and PC relationships.

of $d$ and $d$ is the parent of $e$. The result is sorted lexicographically by $(a, d, e)$.

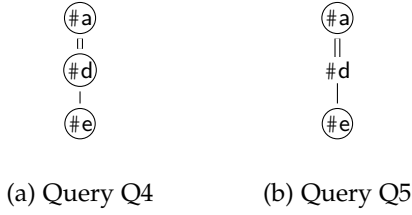

(a) Query Q4    (b) Query Q5

Fig. 4. Sample TPQs

A query plan of Q4 is depicted in Figure 5a. Since all query nodes are core, we utilize only the partial-join algorithms (see Section 6). To produce the sorted result, we need to join the query nodes in a bottom-up manner, so the last join operation produces the result sorted by #a. We first join #d with #e to produce all pairs $[d, e]$ such that $d$ is the parent of $e$. Since we use the ancestor-sorted partial-join (i.e., *StackTreeAnc*), the intermediate result (see Figure 5b) is sorted lexicographically by $(d, e)$. Subsequently, we join the intermediate result with #a to produce all triplets $[a, d, e]$ such that $a$ is an ancestor of $d$; we use *StackTreeAnc* again to make the final result sorted lexicographically by $(a, d, e)$ (see Figure 5c).

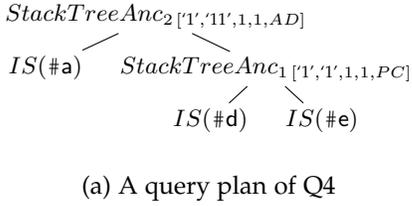

(a) A query plan of Q4

(b) The intermediate result – the result of $StackTreeAnc_1$

(c) The final result – the result of $StackTreeAnc_2$

Fig. 5. An example of the query processing

Now, let us consider the query Q5 in Figure 4b, where #d is a non-output query node. Please note that Figure 5c without the #d column is not the correct result of Q5 since the $[a, e]$ pairs are not sorted properly (see the #e column). Due to this fact, we cannot use the plan of Q4 (Figure 5a) without any change to process Q5. A naive solution is to use the plan in Figure 5a and then to sort the final result by $(a, e)$. However, we lose the advantages of the FP plan since a sorting operation is used (see Section 2.5). To solve this problem without any additional sorting, we introduce an extension of the *StackTreeAnc* algorithm called a *secondary relationship test*: we extend the signature of the algorithm with an attribute $k$.

$$T_o = StackTreeAncSrt_{[M_a, M_d, i, j, \alpha, k]}(T_a, T_d),$$

where $j$ and $k$ refer to the *primary* and the *secondary* join columns in $T_d$, respectively. Let $c_i$ be the join column of $T_a$, let $c_j$ and $c_k$ be the primary and the secondary join columns in $T_d$, respectively. The output is computed such that the input tables are joined by data nodes in $c_i$ and $c_j$ using the traditional *StackTreeAnc* algorithm with the AD relationship, even if $\alpha$ is PC. It requires the table $T_d$ to be sorted by $c_j$. However, the joined tuple is output iff the relationship $\alpha$ is satisfied between data nodes from $c_i$ and $c_k$. This is called the secondary relationship test.

*Example 5.2.* When the novel algorithm is applied, the sorting problem of query Q5 (see Figure 4b) is solved using the plan in Figure 6a. We first join #d with #e using a descendant-sorted partial-join (i.e., the *StackTreeDesc* algorithm), so the pairs $[d, e]$ in the intermediate result are sorted lexicographically by $(e, d)$ (see Figure 6b). Subsequently, we use the novel *StackTreeAncSrt* algorithm joining the intermediate result with data nodes of #a to produce all triplets $[a, d, e]$, such that $a$ is an ancestor of $e$; using the secondary relationship test, we verify whether $a$ is an ancestor of $d$. As a result, the triplets are sorted lexicographically by $(a, e, d)$. Since #d is a non-output query node, we set the first bit of $M_d$ of the *StackTreeAncSrt* to zero to filter $d$ data nodes out.

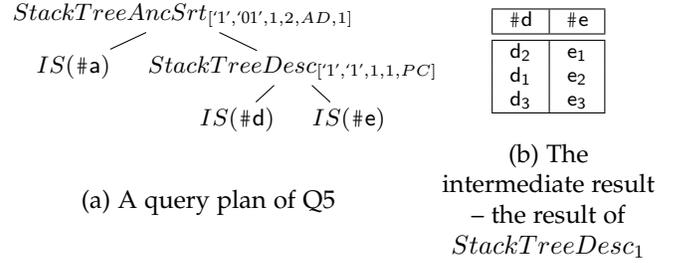

(a) A query plan of Q5

(b) The intermediate result – the result of $StackTreeDesc_1$

Fig. 6. Solving the sorting problem using the novel *StackTreeAncSrt* algorithm

### 5.2 Semi-join Algorithms

In [2], [30], the authors show that the descendant-filtering semi-join for the AD relationship can be implemented without any stack, which leads to a more efficient query processing. In our article, we call this algorithm *SemiJoinDescAD*. To the best of our knowledge, a stack-less ancestor-filtering semi-join for the AD relationship has not yet been presented in the literature; in this section, we present such an algorithm called *SemiJoinAncAD*.

The descendant- and ancestor- filtering semi-joins for the PC relationship are a straightforward simplification of *StackTreeDesc* and *StackTreeAnc* [1] where only the nodes of $T_d$ or $T_a$ form the output. Unfortunately, using of a stack is unavoidable in the case of PC. In this paper, we call the semi-joins *SemiJoinDescPC* and *SemiJoinAncPC*, respectively.

The novel *SemiJoinAncAD* is shown in Algorithm 1. Lines 1 – 9 loop through all nodes in $T_a$ and $T_d$. In Line 3, the stream $T_d$ is advanced since its current data node precedes the current data node of $T_a$. Lines 5 – 6 cover the case when the current node of $T_d$ is a descendant of the current node of $T_a$. Otherwise, the current node of $T_a$ precedes the current node of $T_d$ and the stream $T_a$ needs to be advanced.

**Algorithm 1:** SemiJoinAncAD($T_a$, $T_d$) binary semi-join

    **input** : Two lists $T_a$ and $T_d$ of node labels sorted by the document order.
    **output**: A list of data nodes in $T_a$ having a data node with the AD relationship in $T_d$; the list is sorted by the document order.

**1** **while** $\neg$ finished($T_a$) $\wedge \neg$ finished($T_d$) **do**
**2**   **if** current($T_a$).left $\geq$ current($T_d$).left **then**
**3**     | advance($T_d$);
**4**   **else if** current($T_a$).right $>$ current($T_d$).right **then**
**5**     **output** current($T_a$);
**6**     advance($T_a$);
**7**   **else**
**8**     | advance($T_a$);

## 6 BUILDING FP PLAN

In this section, we show how to combine the proposed partial-join and semi-join algorithms to build an FP plan for a TPQ considering non-output query nodes. An algorithm building an FP plan is presented in [36]. However, this work does not consider non-output query nodes, which is necessary to process XQuery queries. We also show that considering the non-output query nodes means the more efficient processing of queries since semi-joins can be used instead of partial-joins (see Section 8.1). Moreover, sorting by multiple columns has to be considered to get the lexicographically sorted result of an XQuery query in the ordered mode. Our algorithm is called *BuildPlanCore*; it invokes *BuildPlanCons* to build sub-plans for constraining subqueries. We provide examples demonstrating the both algorithms in our technical report [21].

### 6.1 Building FP Plan for Constraining Subquery

The function *BuildPlanCons* building an FP plan for a constraining subquery is shown in Algorithm 2. The algorithm is called with a parameter $q_o$ – a constrained node. In Line 1, a plan $P_{q_o}$ is set to an IS operator. Lines 2 – 16 iterate through all neighbors[2] $q$ of $q_o$ and recursively build a plan $P_q$ (Line 3) for $q$ without the neighbor $q_o$. Then a binary semi-join operator is used to restrict the result of the plan $P_{q_o}$. The result is restricted such that it has to have an ancestor (Line 6), a parent (Lines 8), a descendant (Line 12) or a child (Line 14) in the result of $P_q$. Let us note that the function rel(a, b) returns the XPath relationship between $a$ and $b$.

In Line 2, we iterate through the neighbors in an arbitrary order. In our experiments (see Section 8), we use the order defined by the query syntax. However, it is clear that another order produces another plan. In [36], they try all permutations and select the cheapest one, i.e., a cost-based optimization to select a join order is used. Although this cost-based optimization could be easily integrated in our approach, our experiment (see the technical report [21], Section 8.4) shows that the performance of plans for all permutations do not vary too much. In other words, the performance of any plan is close to the performance of the optimal plan, which is caused by utilizing only the semi-joins instead of the partial-joins. Therefore, in our approach,

---

2. By neighbors of a query node $q$ we mean all its children and its parent (if $q$ is not the root).

**Algorithm 2:** BuildPlanCons($q_o$)

    **input** : Query node $q_o$
    **output**: FP plan $P_{q_o}$ for $q_o$

**1** $P_{q_o} \leftarrow$ IS($q_o$);
**2** **foreach** $q \in$ *neighbors of* $q_o$ **do**
**3**   $P_q \leftarrow$ BuildPlanCons($q$ *without* $q_o$);
**4**   **if** $q =$ parent($q_o$) **then**
**5**     **if** rel($q_o$, $q$) $= AD$ **then**
**6**       | $P_{q_o} \leftarrow$ SemiJoinDescAD($P_q$, $P_{q_o}$);
**7**     **else if** rel($q_o$, $q$) $= PC$ **then**
**8**       | $P_{q_o} \leftarrow$ SemiJoinDescPC($P_q$, $P_{q_o}$);
**9**   **else if** $q_o =$ parent($q$) **then**
**10**     **if** rel($q_o$, $q$) $= AD$ **then**
**11**       | $P_{q_o} \leftarrow$ SemiJoinAncAD($P_{q_o}$, $P_q$);
**12**     **else if** rel($q_o$, $q$) $= PC$ **then**
**13**       | $P_{q_o} \leftarrow$ SemiJoinAncPC($P_{q_o}$, $P_q$);

**14** **return** $P_{q_o}$;

we omit any join order selection which enables us to compare it with holistic joins, where any cost-based optimization is not utilized.

*Example 6.1.*

Let us explain how the algorithm *BuildPlanCons* builds a plan for a sample constraining subquery $\Pi_{\#d}$ in Figure 7a. The algorithm starts with the constrained node #d and it first recursively builds plans for the subtrees of #a and #e without #d itself. Let us refer to these subtrees by $Q_{\#a}$ and $Q_{\#e}$, respectively. The both subtrees are depicted in Figures 7b and 7c. In the recursive call, #a and #e are treated as output query nodes.

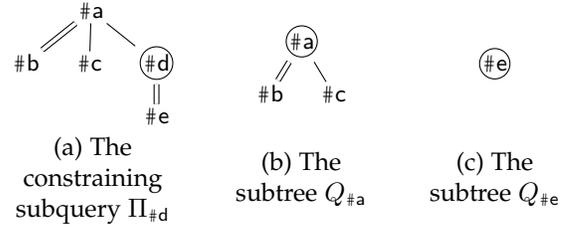

(a) The constraining subquery $\Pi_{\#d}$     (b) The subtree $Q_{\#a}$     (c) The subtree $Q_{\#e}$

Fig. 7. Example of a constraining subquery

The plan for $Q_{\#a}$ is computed such that we first recursively build plans for the subtrees of #b and #c; the plans are $IS(\#b)$ and $IS(\#c)$, respectively, since the subtrees are the query node themselves. The plan for $Q_{\#a}$ is then:

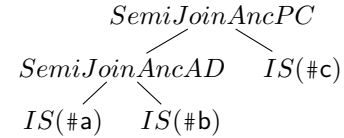

The plan for $Q_{\#e}$ is $IS(\#e)$. Finally, the plan of the constraining subquery $\Pi_{\#d}$ is as follows:

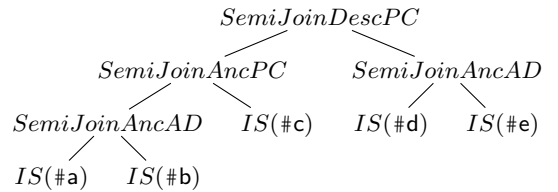

## 6.2 Building FP Plan for Query Core

The function *BuildPlanCore* building an FP plan for a query core is shown in Algorithm 3. The output is a plan producing a table where each column corresponds to one output query node. The order of columns corresponds to the pre-order traversal of the output query nodes. To meet the XQuery semantics, the table is lexicographically sorted by the columns.

---

**Algorithm 3:** BuildPlanCore($q$)

**input** : Query node $q$
**output**: FP plan $P$ for $q$

1  $P \leftarrow$ BuildPlanCons($\Pi_q$);
2  **foreach** $c \in$ core_children($q$) **do**
3  $\quad \alpha \leftarrow$ rel($q$, $c$);
4  $\quad$ **if** isOutput($c$) **then**
5  $\quad\quad R \leftarrow$ BuildPlanCore($c$);
6  $\quad\quad P \leftarrow StackTreeAnc_{['1*','1*',1,1,\alpha]}(P, R)$;
7  $\quad$ **else**
8  $\quad\quad R \leftarrow$ BuildPlanCons($\Pi_c$);
9  $\quad\quad i \leftarrow 1$;
10 $\quad\quad d \leftarrow c$;
11 $\quad\quad$ **while** $\neg$ isOutput($d$) **do**
12 $\quad\quad\quad d \leftarrow$ core_child($d$);
13 $\quad\quad\quad$ **if** $\neg$ isOutput($d$) **then**
14 $\quad\quad\quad\quad S \leftarrow$ BuildPlanCons($\Pi_d$));
15 $\quad\quad\quad$ **else**
16 $\quad\quad\quad\quad S \leftarrow$ BuildPlanCore($d$);
17 $\quad\quad\quad \beta \leftarrow$ rel($d$, parent($d$));
18 $\quad\quad\quad R \leftarrow StackTreeDesc_{['10*','1*',i,1,\beta]}(R, S)$;
19 $\quad\quad\quad i \leftarrow 2$;
20 $\quad\quad P \leftarrow StackTreeAncSrt_{['1*','01*',1,2,\alpha,1]}(P, R)$;

21 **return** $P$;

---

The algorithm is invoked with the root query node of the query core as a parameter. In Line 1, it builds a plan $P$ for a constraining subquery of the root query node of the query core. In Lines 2 – 20, the algorithm loops through all child query core nodes $c$ of $q$ and successively extends the plan $P$. The function core_children($q$) returns all query core nodes that are children of $q$.

If $c$ is an output query node, a plan $R$ for $c$ is recursively built (Line 5) and it is joined with $P$ using *StackTreeAnc* (Line 6). A more complicated situation becomes if $c$ is a non-output query node. A plan $R$ for the constraining subquery of $c$ is built (Line 8) and the child query core node of $c$ is assigned to $d$ (Line 12). The function core_child($q$) returns the first child query core node of $q$. It is guaranteed that if $q$ is a non-output query core node, it has exactly one child query core node (we discuss this property in Appendix B). Now, if $d$ is still a non-output query node, a plan $S$ is built using *BuildPlanCons* (Line 14). Otherwise, *BuildPlanCore* is recursively invoked (Line 16). In Line 18, the plans $R$ and $S$ are joined using *StackTreeDesc*. According to the variable $i$, the first or the second column of $R$ is used as the join column.

We repeat the navigation to the non-output query core nodes (Lines 11 – 19) until an output query node is found. In Line 20, $R$ and $P$ are joined using *StackTreeAncSrt*. The secondary relationship test (see Section 5.1) with the first column of $R$ ($k = 1$) ensures that the output satisfies the required relationship $\alpha$.

Unlike from *BuildPlanCons*, *BuildPlanCore* has to iterate through child query nodes (Lines 2 – 20) in the order defined by the query syntax, so the result is sorted by output query nodes in the pre-order traversal (see Section 2.2). In [36], they compute the cost of all orders and pick the cheapest one which, however, means that the lexicographically sorted result is not considered. Therefore, our algorithm is deterministic and, for a query core, it generates exactly one FP plan; it is not necessary to utilize any cost-based optimization technique.

*Example 6.2.*

Let us demonstrate how the algorithm *BuildPlanCore* builds a plan for the query core in Figure 2a. The algorithm starts with the parameter $q = $ #a, since #a is the root of the query core. The first step is to compute a plan $P_{\#a}$ for the constraining subquery $\Pi_{\#a}$ (Line 1).

After that we loop through the children of #a. #b is the first one of them and it is a non-output query node. This is the more complicated case handled by Lines 8 – 20 in Algorithm 3. We start with the building of a plan $R$ for the constraining subquery of #b (Line 8). Since the constraining subquery of #b is the query node itself, the plan $R = IS(\#b)$. In Line 12, $d$ is set to #c. Since #c is an output query node, we recursively invoke *BuildPlanCore* to compute a plan $S$. Since #c does not have any children, the plan $S = IS(\#c)$. In Line 18, we join the plans $R$ and $S$ using *StackTreeDesc*, the plan $R$ is now:

$$StackTreeDesc_{['1','1',1,1,AD]}$$
$$IS(\#b) \quad IS(\#c)$$

Such a plan produces a table of two columns for the query nodes #b and #c ordered by the second column, respectively. Since the variable $d$ is now an output query node, the loop in Line 11 is terminated and we move to Line 20; we join $P$ with $R$ using *StackTreeAncSrt*:

$$StackTreeAncSrt_{['1','01',1,2,PC,1]}$$
$$P_{\#a} \quad StackTreeDesc_{['1','1',1,1,AD]}$$
$$IS(\#b) \quad IS(\#c)$$

The mask $M_d$ the *StackTreeAncSrt* is set to '01' since we want to retain only the second column corresponding to #c computed by $R$, because #b is a non-output query node. The indexes $i$ and $j$ are set to 1 and 2, respectively, since we need to join data nodes of #a with the data nodes of #c. We use the secondary relationship test to ensure that the data nodes of #b are in the PC relationship with #a.

Finally, the plan $P$ is extended by the plan $P_{\#d}$ of the constraining subquery for #d. The result of the algorithm is the following plan:

$$StackTreeAnc_{['11','1',1,1,AD]}$$
$$StackTreeAncSrt_{['1','01',1,2,PC,1]} \quad P_{\#d}$$
$$P_{\#a} \quad StackTreeDesc_{['1','1',1,1,AD]}$$
$$IS(\#b) \quad IS(\#c)$$

10## 7 ANALYSIS OF THE BINARY JOIN APPROACH

For the purposes of our analysis, let us define the following properties of a stream. A *recursive stream* is a stream with some data nodes having the AD relationship. Recursive streams are caused by recursive tags in an XML document and they have a significant impact on the algorithm complexity [4]. A *repeating stream* is a stream with repeating data nodes, i.e., a repeating stream can return a particular data node multiple times. A repeating stream can be produced by a partial-join. For example, when the intermediate result in Figure 5c is an input of another join and #a is the join column then the intermediate result is a repeating stream. A *simple stream* is a non-recursive and non-repeating stream.

The cornerstone of every theoretical analysis of a TPQ processing approach [1], [7], [8], [9], [18], [20], [17], [4] is the following Definition 1.

*Definition 1.* (Optimal TPQ processing approach) If an approach is optimal for a TPQ, then the time complexity is linear with respect to the size of the input and output and the space complexity is linear with respect to $d$, where $d$ is the depth of an XML document.

Until now the optimality has been proven only for holistic join approaches. Their optimality is proven with respect to the TPQ properties [7], [9] (e.g., all holistic approaches are optimal for a TPQ having only AD relationships) or with respect to the recursiveness of input streams [4]. We prove the optimality of our binary join approach in Section 7.1 and we compare it with holistic join approaches in Section 7.2.

### 7.1 Binary Join Approach Optimality

*Definition 2.* (The linear time complexity constraint). A binary join algorithm with $T_a$ and $T_d$ input and $T_o$ output streams of node labels sorted by the document order satisfies the linear time complexity constraint if $|T_o| \leq |T_a| + |T_d|$, where $|T_o|$ is the size of the output and $|T_a|$ and $|T_d|$ are sizes of the inputs.

*Lemma 7.1.* (The linear time complexity of the binary join approach for a TPQ $Q$). Let $\sum |T_i|$ be the sum of sizes of all input streams defined by query nodes in $Q$. Our binary join approach processing $Q$ has the linear time complexity with respect to $\sum |T_i|$ if each binary join in the corresponding FP plan satisfies the linear time complexity constraint.

*Proof:* If each binary join in an FP plan satisfies the linear time complexity constraint, then, in the worst case, for each binary join in the FP plan $|T_o| = |T_a| + |T_d|$. In such a case, for the root binary join $|T_a| + |T_d| = \sum |T_i|$. For each other binary join in the plan $|T_a| + |T_d| < \sum |T_i|$. Therefore, since (1) for each binary join algorithm in an FP plan $|T_a| + |T_d| \leq \sum |T_i| \wedge |T_o| \leq \sum |T_i|$ and (2) a binary join algorithm works with the linear time complexity with respect to the size of its input and output [1], the binary join approach has the linear time complexity with respect to $\sum |T_i|$. □

The linear time complexity constraint is naturally satisfied for all semi-joins since for *SemiJoinAncAD* and *SemiJoinAncPC* $|T_o| \leq |T_a|$ and for *SemiJoinDescAD* and *SemiJoinDescPC* $|T_o| \leq |T_d|$. It is also satisfied for the partial-joins *StackTreeDesc* and *StackTreeAnc* when $T_a$ is simple since each data node from $T_d$ is joined with at most one data node from $T_a$.

*Definition 3.* (The linear space complexity constraint). A binary join algorithm satisfies the linear space complexity constraint if it has the linear space complexity with respect to $d$.

*Lemma 7.2.* (The space complexity of the binary join approach for a TPQ $Q$). Our binary join approach has the linear space complexity with respect to $d$ if each binary join algorithm in an FP plan corresponding to $Q$ satisfies the linear space complexity constraint.

*Proof:* If each binary join algorithm in an FP plan satisfies the linear space complexity constraint, then each binary join works with the linear space complexity with respect to $d$. Since $d$ is the same for each algorithm in the FP plan, the binary join approach works with the linear space complexity with respect to $d$. □

Let us discuss which binary join algorithms satisfy the linear space complexity constraint. We start with the semi-joins *SemiJoinAncAD* and *SemiJoinDescAD*. Since these semi-joins do not use any stack or lists (see Algorithm 1) they have the constant space complexity. The space complexity of other binary joins is affected mainly by the type of the stream $T_a$:

- If $T_a$ is simple, there is at most one entry in the stack of partial-joins *StackTreeDesc* and *StackTreeAnc*. For *StackTreeAnc*, it is also not necessary to use the self- or inherited-lists [1]. Therefore, in this case, the space complexity of the partial-joins is constant. The same property is valid for semi-joins *SemiJoinAncPC* and *SemiJoinDescPC* since they are derived from the partial-joins.
- If $T_a$ is recursive, the partial-join *StackTreeDesc* and the semi-join *SemiJoinDescPC* can store more data nodes from $T_a$ on the stack, however there are no self- or inherited-lists and the size of the stack is bounded by $d$. The partial-join *StackTreeAnc* and the semi-join *SemiJoinAncPC* have to use the self- and inherited- lists and the size of these lists is not bounded by $d$. We discuss an example of such a situation in Appendix C.
- If $T_a$ is repeating, the size of the stack of the partial-joins *StackTreeDesc* and *StackTreeAnc* is not bounded by $d$.

Table 3 summarizes which binary join algorithms satisfy the linear time complexity constraint (TC) (Definition 2) and the linear space complexity constraint (SC) (Definition 3) with respect to the properties of $T_a$. Following Section 6, all semi-joins in an FP plan are always processed before all partial-joins. Since only the partial-joins can produce repeating streams, we do not analyse properties of semi-joins if $T_a$ is repeating. We also do not explicitly analyse properties of the novel *StackTreeAncSrt* (see Section 5.1) since it has the same properties as *StackTreeAnc*. Using the Table 3 and Lemmas 7.1 and 7.2, we can now derive several classes of queries for which our binary join approach is optimal.

| Algorithm | | $T_a$ is | | | | | |
|---|---|---|---|---|---|---|---|
| | | simple | | recursive | | repeating | |
| | | TC | SC | TC | SC | TC | SC |
| semi-joins | *SemiJoinDescAD* | ✓ | ✓ | ✓ | ✓ | – | – |
| | *SemiJoinAncAD* | ✓ | ✓ | ✓ | ✓ | – | – |
| | *SemiJoinDescPC* | ✓ | ✓ | ✓ | ✓ | – | – |
| | *SemiJoinAncPC* | ✓ | ✓ | ✓ | ✗ | – | – |
| part.-joins | *StackTreeDesc* | ✓ | ✓ | ✗ | ✓ | ✗ | ✗ |
| | *StackTreeAnc* | ✓ | ✓ | ✗ | ✗ | ✗ | ✗ |

TABLE 3
Linear time and space complexity constraints of the binary join algorithms for $T_a$

*Lemma 7.3.* (Optimal processing of constraining subquery). The binary join approach is optimal to process a constraining subquery if there is not an edge with the following properties: (i) having the PC relationship, and (ii) not lying on a path from the root to the constrained node, and (iii) having the recursive stream of the parent query node.

*Proof:* A constraining subquery is processed only by the semi-join algorithms (see Algorithm 2). An edge with the PC relationship not lying on a path between the root and the constrained node leads to the usage of *SemiJoinAncPC*. This is the only semi-join algorithm not satisfying the linear space complexity constraint if $T_a$ is recursive. □

*Theorem 1.* (TPQ with one output query node). The binary join approach is optimal to process a TPQ with one output query node if there is not an edge: (i) with the PC relationship, and (ii) not lying on a path from the root to the output query node, and (iii) having the recursive stream of the parent query node.

*Proof:* Theorem 1 directly follows Lemma 7.3 since each TPQ with one output query node is a constraining subquery itself. □

*Theorem 2.* (TPQ with more output query nodes). The binary join approach is optimal for a TPQ with more output query nodes if (i) it is optimal for all constraining subqueries, and (ii) the query core is a path, and (iii) the streams of the query core nodes are not recursive.

*Proof:* Let us first prove Theorem 2 for a query where all query core nodes are output. Each edge in the query core is then processed by *StackTreeAnc* (see Algorithm 3). If the query core is a path, $T_a$ for each *StackTreeAnc* is computed by semi-joins of a constraining subquery, therefore $T_a$ is not repeating. Only $T_d$ can be computed by another *StackTreeAnc* and, therefore, it can be repeating. Since $T_a$ is not repeating and not recursive, it is simple.

If there are non-output query nodes, each edge in the query core, where the parent is a non-output query node, is processed by *StackTreeDesc* (see Algorithm 3). The output stream $T_o$ of *StackTreeDesc*, where $T_a$ and $T_d$ are simple, is also simple since a data node from $T_d$ can be joined with at most one data node from $T_a$. Since $T_a$ of *StackTreeDesc* is the output of a constraining subquery or of another *StackTreeDesc*, $T_a$ is simple. If $T_a$ is simple for each partial-join in an FP plan, the binary join approach is optimal. □

*Example 7.1.*
In Figure 8a, there is a query Q6 without non-output query nodes. The whole query is the query core itself and it is a path. The corresponding FP plan is depicted in Figure 8b. Column names (corresponding to query nodes) of intermediate results are depicted in angle brackets. Let us consider that the streams of all query nodes are non-recursive. In such a case, the inputs $T_a$ of all *StackTreeAnc* partial-joins are simple. Therefore, the binary approach processing the FP plan for Q6 is optimal.

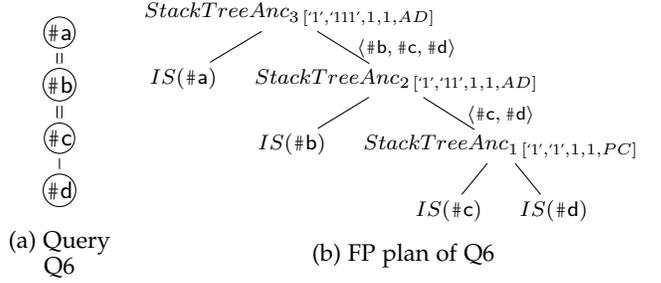

(a) Query Q6     (b) FP plan of Q6

Fig. 8. A query core without non-output query nodes

*Example 7.2.*
The query Q7 in Figure 9a is also a query core itself and it is a path, however, #b and #c are non-output query nodes. The corresponding FP plan is depicted in Figure 9b. $T_a$ of $StackTreeDesc_2$ is the output of $StackTreeDesc_1$. Assuming that the streams for #b and #c are non-recursive, each data node of #c is joined with at most one data node of #b. Therefore the output of $StackTreeDesc_1$ is non-recursive and non-repeating, i.e., it is simple. Since $T_a$ of all partial-joins are simple, the binary approach processing the FP plan for Q7 is optimal.

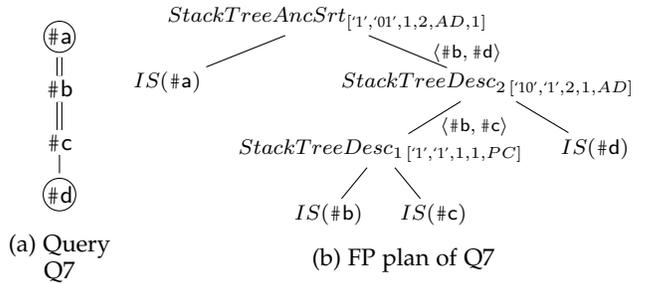

(a) Query Q7     (b) FP plan of Q7

Fig. 9. A query core with non-output query nodes

## 7.2 Comparison with Holistic Approaches

Let us compare the optimality of our binary join approach and the holistic join approaches. Every holistic join is optimal when a query contains only AD relationships [7]. A holistic join approach is optimal also for a TPQ if all streams corresponding to the query nodes having the PC relationship are non-recursive [4]. When compared to the novel optimality of our binary join approach, *the only difference is that the optimality of holistic approaches is not influenced by the number of non-output query nodes.*

## 8 EXPERIMENTAL RESULTS

In this section, we experimentally demonstrate weak and strong aspects of binary and holistic join approaches under certain conditions. Both holistic join (HJ) and binary join (BJ) approaches are implemented in our native database management system called RadegastXDB (RXDB) [11]. Both approaches use the same index data structure (see Section 2.7) and no cost-based optimizer to select a query plan is included. Let us recall that we select the *GTPStack* holistic join [4] as a representative of the state-of-the-art holistic algorithms.

We employed three typical data collections with various characteristics and sizes, these collections are commonly used in many approaches [24], [10], [17], [4]. XMark [29] is a well-known XML benchmark representing a synthetic database of auctions. An XML document of auctions can be generated with the factor $f$ that controls the size of the document; we used two factors: $f = 1$ and $f = 10$. The structure of the document is rather regular with a few recursive tags. SwissProt [32] is a real database of protein sequences with annotations; its structure is quite irregular without any recursive tags. TreeBank [32] is a real database of partially encrypted English sentences; it is characterized by a highly recursive and irregular structure. TreeBank is the smallest collection, but it is the most complicated collection from the query processing perspective due to its highly recursive nature. An overview of the most important statistics of these collections can be found in Table 4. We also show the size of the index data structures in RXDB.

| Collection | Size (MB) | Number of data nodes | Max depth | Index size (MB) |
|---|---|---|---|---|
| XMark (f=1) | 111 | 2,048,193 | 14 | 31 |
| XMark (f=10) | 1,137 | 20,532,805 | 14 | 315 |
| SwissProt | 109 | 5,166,890 | 7 | 79 |
| TreeBank | 82 | 2,437,667 | 38 | 37 |

TABLE 4
Statistics of data collections

For each data collection, we generated 60 unique *twig patterns* combining the AD and PC relationships. By a twig pattern we mean a TPQ without a specification of non-output and output query nodes. Each twig pattern has from 3 to 12 query nodes, its depth is from 3 to 9 and each query node has 3 chidren at maximum. The complete list of the twig patterns can be found in Appendix E. For each twig pattern $p$, we randomly generated $n$ queries where $n$ is the number of query nodes in $p$, such that each query has a different number of output query nodes. Finally, there are 356 queries for XMark, 334 queries for SwissProt, and 393 queries for TreeBank. The twig patterns were generated such that each setting of output or non-output query nodes produces a non-empty result, i.e., the *selectivity* is greater than zero. We compute the selectivity of a query as $\sigma = \frac{N_{out}}{N_{in}}$, where $N_{out}$ is the total number of distinct data nodes in the result of the query and $N_{in}$ is the total number of data nodes in streams specified by output query nodes.

The experiments were performed on a machine with Intel Xeon E5-2690@2.9 GHz processor and the Microsoft Windows Server 2008 R2 Datacenter (SP1) operating system. During the experiments, the index data structures of RXDB were completely loaded in the main memory.

### 8.1 Non-output and Output Query Nodes

This experiment shows how the number of non-output and output query nodes affects the processing time of HJ and BJ. A typical example is demonstrated in Figure 10. For the demonstration, we used the twig pattern TB18 with 7 query nodes, therefore, 7 queries with the different number of output query nodes have been tested.

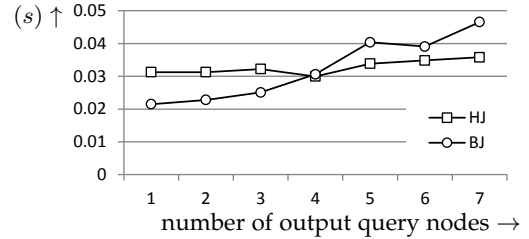

Fig. 10. Comparison of HJ and BJ on a sample query (TB18)

We can see that the processing time of BJ is more influenced by the number of output query nodes compared to the processing time of HJ. For a query with one output query node, BJ uses only semi-joins and it outperforms HJ. On the other hand, when all query nodes are set to output, only partial-joins are used and HJ outperforms BJ.

Summarized results for all collections can be seen in Figure 11. Horizontal axis represents the ratio $\rho = \frac{n_o}{n}$, where $n_o$ is the number of output query nodes and $n$ is the total number of query nodes in a query. The values on the vertical axis represent the relative speed-up of BJ compared to HJ; it is computed as $\frac{t_{HJ}}{t_{BJ}}$, where $t_{HJ}$ and $t_{BJ}$ are total processing times of queries with $\rho$ in the specific range using HJ and BJ, respectively, i.e., BJ outperforms HJ when $\frac{t_{HJ}}{t_{BJ}} > 1$. We can see that the lower value of $\rho$ means the lower processing time of BJ compared to HJ. This is consistent with the analysis in Section 7: the more output query nodes there are in a query, the more repeating streams appear in the corresponding FP plan, and therefore, BJ becomes non-optimal. On the other hand, the optimality of HJ does not depend on the number of output query nodes. We can see that for TreeBank, when $\rho \in (.8 - 1]$, HJ outperforms BJ, which is caused by the highly recursive nature of the TreeBank collection. However, the speed-up is relatively small. We can also observe that the results for both XMark collections are very similar, which means that the relative speed-up of BJ does not depend on the size of the collection.

Figure 11 also compares our BJ with another BJ presented in [36], where all query nodes are always output. The relative speed up of [36] to HJ can be therefore approximated by $\frac{t_{HJ}}{t_{BJ}}$ where $\rho \in (.8; 1]$. For such a range of $\rho$, the performance of BJ is the worst. We can see that considering the non-output query nodes means a more efficient query processing in the case of BJ.

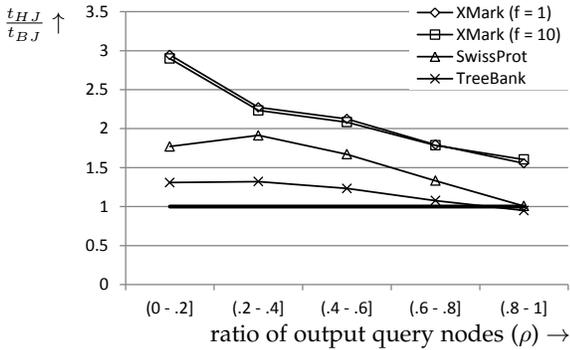

Fig. 11. Comparison of HJ and BJ on queries with various ratios of output query nodes

total number of data nodes in the result[3]. The last column compares the intermediate result size to the final result size.

| Collection | $\max(\mu)$ | Result size | $n_o$ | $\frac{\max(\mu)}{n_o \times [\text{result size}]}$ (%) |
|---|---|---|---|---|
| one output query node ||||| 
| XMark (f = 1) | 12 | 232 | 1 | 5.17 |
| XMark (f = 10) | 13 | 2,374 | 1 | 0.55 |
| SwissProt | 7 | 11,099 | 1 | 0.06 |
| TreeBank | 59 | 20 | 1 | 295.00 |
| all output query nodes ||||| 
| XMark (f = 1) | 902 | 111,676 | 6 | 0.13 |
| XMark (f = 10) | 1,280 | 1,131,354 | 6 | 0.02 |
| SwissProt | 79,341 | 149,691 | 7 | 7.57 |
| TreeBank | 11,696 | 307 | 10 | 380.98 |

TABLE 5
Maximum intermediate storage sizes

## 8.2 Intermediate Result Sizes

In this experiment, we focus on the size of the intermediate storage of the binary join approach, since it is often identified as a weak aspect of binary join approaches in the literature. By the intermediate storage we mean all dynamic data structures used by binary joins in an FP plan: stacks, inherited-lists, and self-lists. For each query, we measured $\mu$ which represents the maximum number of data nodes in the intermediate storage.

Summarized results are shown in Figure 12. For five ranges of $\rho$ in the horizontal axis, we computed arithmetic means of $\mu$ for each data collection separately. We can observe that for a small $\rho \in (0;.2]$, in the worst case, there are approximately 10 data nodes in the intermediate storage at one time of the query processing. For $\rho \in (.8;1]$, the value of $\overline{\mu}$ is approximately 100 – 5,000. The higher ratio of output query nodes $\rho$ means using of partial-joins in the FP plan, which causes the repeating streams that have a negative impact on the space complexity (see Section 7). The results show that the considering of non-output query nodes significantly reduces the intermediate storage size.

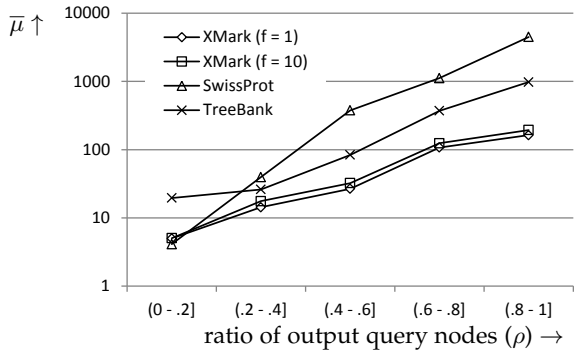

Fig. 12. Maximum intermediate storage size for queries with variable ratio of output query nodes

In Table 5, we show $\max(\mu)$ of queries with one and all output query nodes. For the queries where $\mu = \max(\mu)$, we include the result size and the number of output query nodes $n_o$. In the last column, we show a ratio of $\max(\mu)$ to $n_o \times [\text{result size}]$, where $n_o \times [\text{result size}]$ means the

For the queries with one output query node, we can see that the intermediate storage includes rather small numbers of data nodes. In the worst case, $\max(\mu) = 59$ (for TreeBank), it is approximately $3\times$ more than the number of data nodes in the result. Let us note that the number of data nodes in the result does not represent the byte size of the result. In this paper, the node label consists of 3 integer numbers, i.e., 12 B. In such a case, the byte size of the intermediate storage is never higher than $59 \times 12 = 708$ B.

For the queries where all query nodes are output, $\max(\mu)$ is significantly higher than in the case of queries with one output query node. For SwissProt, $\max(\mu) = 79,341$ (approximately 1 MB of memory), and for TreeBank, $\max(\mu) = 11,696$ which is almost $4\times$ more than the number of data nodes in the result. Such a behavior shows a typical disadvantage of a binary join approach not considering non-output query nodes: the large intermediate result can be produced even if the final result is rather small. However, we can see that this behavior is observed only for one data collection in the experiment. Moreover, when 12 B as the label size is considered the worst case memory size necessary during the query processing is less then 1 MB which can be easily accommodated in the L2 cache of CPU.

## 8.3 Query Selectivity

This experiment is focused on the processing time of BJ and HJ on queries with different selectivity. The results are presented in Figure 13, where the horizontal axis represents specific ranges of the selectivity and the values on the vertical axis represent the relative speed-up of BJ compared to HJ; it is computed again as $\frac{t_{HJ}}{t_{BJ}}$.

As a result, the lower the selectivity $\sigma$ is, the more HJ outperforms BJ. On all data collections, HJ performs better when $\sigma < 0.01$. On the other hand, for higher values of $\sigma$, BJ outperforms HJ. For example, on the XMark collection (both factors), for queries where $\sigma$ is close 1, BJ is approximately $3\times$ faster than HJ. HJ performs better on high-selective queries, since it uses a more aggressive forwarding of the input streams. For low-selective queries, BJ gives better results since it utilizes more simple logic to advance the input streams.

3. The result size of a query is the total number of output matches (tuples). To retrieve the total number of data nodes in the result, the result size has to be multiplied by the number of output query nodes.

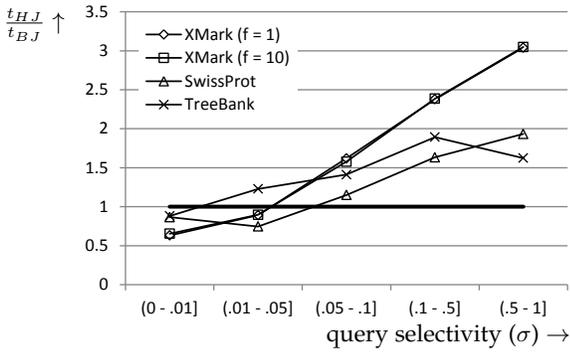

Fig. 13. Comparison of HJ and BJ for queries with different selectivity

We repeated the same experiment focusing only on queries where all query nodes are output. Let us recall that in such a case, the queries are query cores themselves and, therefore, no semi-joins are included in their query plans. The results are depicted in Figure 14. Even though the results of BJ are worse than in Figure 13, we can see that BJ still outperforms HJ on queries with $\sigma > 0.05$. Such an experiment shows that the binary join approach outperforms holistic approaches for low-selective queries even if we consider the traditional TPQ model without a specification of non-output query nodes.

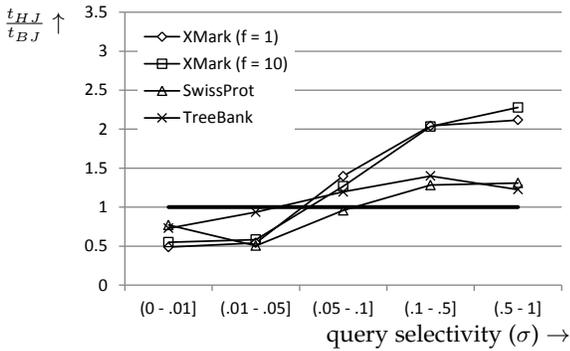

Fig. 14. Comparison of HJ and BJ for queries with different selectivity and all output query nodes

### 8.4 Query Plans

In Section 6, we proposed the algorithm *BuildPlanCons* to build an FP plan for a constraining subquery. The algorithm can iterate through children of a query node in an arbitrary order and, therefore, it can generate a number of FP plans. This experiment shows that the processing times of different FP plans for a query do not vary too much. We tested queries with one output node for TreeBank where at least 12 different FP plans could be generated.

The results are depicted in Figure 15 with box plots of the processing times of BJ using different FP plans (the processing times of HJ are also included). We can see that the processing times of different plans vary within a few (less than 15) milliseconds. As we can observe in most of the queries, the whole box plots are either above or bellow the time of HJ, which means that the processing time of BJ is usually either better or worse than the processing time of HJ, no matter what plan is used. We repeated the same experiment over the other collections (XMark and SwissProt) with similar results. The results show that it is possible to use *BuildPlanCons* even without any cost-based optimizer to select an order of the semi-joins. On the other hand, a cost-based optimizer (e.g., the join order selection proposed in [36]) can be simply integrated in our approach.

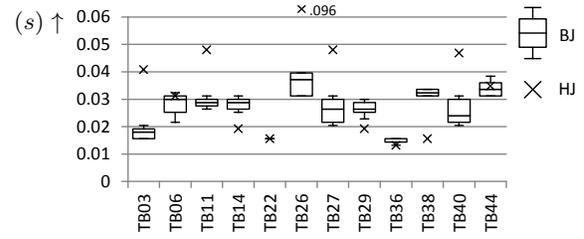

Fig. 15. Processing different query plans

### 8.5 Combined and Cost-Based Approach

Following the experimental results in Section 8.1 and Section 8.3, let us propose two approaches taking into account advantages of the both BJ and HJ.

#### 8.5.1 Combined Join Approach

The results in Section 8.1 clearly show that BJ outperforms HJ when there is a low ratio of output query nodes in a TPQ. Based on this observation, we propose a combined join approach (CJ), where all constraining subqueries are evaluated using binary semi-joins in an FP plan and the query core is evaluated by a holistic join. That means, in extreme cases: if there is one output query node in a query, CJ processes the query in the same way as BJ, and if all query nodes are output CJ processes the query in the same way as HJ. The total processing times of HJ, BJ, and CJ approaches on all testing queries for the collections are depicted in Table 6.

| Collection | HJ | BJ | CJ | Best | CB-J |
|---|---|---|---|---|---|
| XMark (f = 1) | 4.806 | 2.486 | 3.727 | 2.156 | 2.290 |
| XMark (f = 10) | 49.343 | 25.455 | 38.665 | 22.586 | 23.622 |
| SwissProt | 14.561 | 10.718 | 12.031 | 9.448 | 9.916 |
| TreeBank | 12.659 | 11.286 | 10.975 | 9.347 | 10.037 |
| Total | 81.369 | 49.946 | 65.398 | 43.537 | 45.865 |

TABLE 6
Comparison of BH, HJ, CJ

We can observe that CJ outperforms HJ, but for the collections XMark (f=1, f=10) and SwissProt it is worse than BJ. That is because even if all nodes in a query are output, BJ can still outperform HJ (see Figure 14).

#### 8.5.2 Cost-based Approach

Following the previous experiment, to choose an appropriate approach (HJ, BJ or CJ) for a query, we need to take the query selectivity into account. The results of the experiments in Section 8.3 show that BJ outperforms HJ if the query selectivity is not too high ($\sigma$ is not a small value). Therefore, we propose a cost-based approach (CB-J) based on a simple cost model:

**if** $\sigma < \Sigma$: use **HJ**   **if** $\sigma_{core} > \Sigma_{core}$: use **BJ**   **else**: use **CJ**

where $\sigma_{core}$ represents the selectivity of the query core and $\Sigma$ and $\Sigma_{core}$ are constants (thresholds). The decision rule is interpreted such that if the selectivity $\sigma$ of the query is high, it is appropriate to process the whole query using HJ. Otherwise, we check $\sigma_{core}$ to decide whether to use BJ or CJ. Let us note that the query selectivity can be estimated (e.g., [26], [22]), but the estimation is beyond the scope of this paper.

The total query processing times of CB-J are also depicted in Table 6[4]. The table also includes the value *Best* which means the total processing time when the best of the three approaches (BJ, HJ, CJ) is used for each query; i.e., it represents the processing time of the ideal (100 % accurate) cost model. We can see that for all collections, CB-J outperforms HJ and BJ and it is close to the best possible processing time.

There are far more options how to combine HJ a BJ and, therefore, a more robust cost model should be established. For example, it is possible to employ a plan selection method similarly as in [36] or [33]. However, this is beyond the scope of this paper and it is a motivation for our future work.

### 8.6 Comparison of Native XML DBMSs

Finally, we compare our novel algorithms implemented in RXDB with other commonly used systems: BaseX[5] (BX) and MonetDB with an XQuery front-end[6] (MDB). Summarized results are shown in Table 7: the values represent the total processing times of all testing queries in seconds. We can see that the best overall performance on all data collections is given by our RXDB using the binary join approach (R-BJ); R-BJ is approximately $2 - 5\times$ faster than MDB and by one order of magnitude faster than BX. R-HJ (RXDB using the holistic join approach) gives worse overall performance compared to R-BJ, but it can process particular queries with a high ratio of output query nodes or highly-selective queries more efficiently (see Sections 8.1 and 8.3, respectively).

| Collection | BX | MDB | R-BJ | R-HJ |
|---|---|---|---|---|
| XMark (f=1) | 27.7 | 15.2 | **3.2** | 6.3 |
| XMark (f=10) | 280.1 | 62.0 | **33.3** | 64.6 |
| SwissProt | 206.6 | 57.4 | **12.7** | 15.9 |
| TreeBank | 129.7 | 29.7 | **11.2** | 11.6 |

TABLE 7
Total query processing times (s)

Table 8 shows processing times of queries processed on XMark ($f = 10$) and TreeBank listed in Appendix F. The first 5 queries on both collections have one output query node, the other 5 queries have all output query nodes. The lowest time for each query is highlighted. We can see that the relative speed-up of R-BJ and R-HJ over MDB or BX is similar for all the queries. However, for queries with all output nodes on TreeBank, R-HJ gives lower processing times than R-BJ, which supports our analysis in Section 7.

---

4. We experimentally picked the constants $\Sigma = 0.001$, $\Sigma_{core} = 0.1$.
5. http://basex.org/
6. https://www.monetdb.org/XQuery/

| | XMark (f=10) | | | | TreeBank | | | |
|---|---|---|---|---|---|---|---|---|
| | BX | MDB | R-BJ | R-HJ | BX | MDB | R-BJ | R-HJ |
| 1 | **.023** | .037 | **.023** | **.023** | .285 | .062 | **.012** | .016 |
| 2 | .284 | .131 | **.031** | .109 | .499 | .109 | **.023** | .043 |
| 3 | **.030** | .059 | .060 | .057 | .093 | .051 | **.008** | .014 |
| 4 | .736 | .351 | **.062** | .224 | .306 | .109 | **.031** | .031 |
| 5 | .376 | .183 | **.047** | .220 | .251 | .069 | **.031** | .057 |
| 6 | 1.218 | .351 | **.255** | .310 | .622 | .118 | **.094** | .098 |
| 7 | 1.345 | **.126** | .127 | .199 | .618 | .110 | .031 | **.021** |
| 8 | .874 | .167 | **.092** | .197 | .397 | .071 | .021 | **.020** |
| 9 | .463 | .059 | **.047** | .107 | 1.117 | .281 | .045 | **.027** |
| 10 | 1.263 | .346 | **.242** | .472 | .567 | .101 | .059 | **.047** |

TABLE 8
Processing times of particular queries (s)

## 9 CONCLUSION

In this article, we introduced a new binary join approach processing a TPQ with defined output and non-output query nodes using an FP plan (i.e., the plan without an intermediate result materialization and explicit sorting); considering output and non-output query nodes supports the XQuery semantics which leads to a more efficient query processing in many cases. We also showed that it is not necessary to integrate a cost-based optimizer to select a query plan in this approach since the performance of all individual plans is rather similar. Moreover, we proved that for a certain class of TPQs such an FP plan has the linear time complexity with respect to the size of the input and output and the linear space complexity with respect to the XML document depth (i.e., the same complexity as the holistic join approaches). We experimentally compared the binary approach with a state-of-the-art holistic approach (*GTPStack*) for 4 data collections with different characteristics: although we showed that our binary approach overcomes *GTPStack* when the average processing time is considered, we put forward that *GTPStack* performs more efficiently for the high-selective queries for all data collections (up-to the selectivity 0.05) and for the queries with a lower ratio of non-output query nodes for two data collections. As a result we propose a simple approach combining advantages of both types of approaches using a simple heuristic.

We also compared the binary approach with another FP plan-enabling binary approach and with other commonly used XML database management systems. In the case of the comparison with another binary approach, we mainly showed that the consideration of non-output query nodes decreases the query processing time up-to $2.5\times$. When compared with the other XML database management systems, our approach lowers the average query processing time from $2\times$ to $17\times$. In the experiments, we also showed that the intermediate result sizes of our binary approach did not exceed 1 MB in the worst case although the high intermediate result sizes of binary join approaches are often criticized in the literature.

## REFERENCES

[1] S. Al-Khalifa, H. Agadish, N. Koudas, J. M. Patel, D. Srivastava, and Y. Wu. Structural joins: A primitive for efficient xml query pattern matching. In *Data Engineering, 2002. Proceedings. 18th International Conference on*, pages 141–152. IEEE, 2002.
[2] S. Al-Khalifa and H. Jagadish. Multi-level operator combination in xml query processing. In *Proceedings of the eleventh international conference on Information and knowledge management*, pages 134–141. ACM, 2002.

[3] R. Bača, M. Krátký, I. Holubová, M. Nečaský, T. Skopal, M. Svoboda, and S. Sakr. Structural xml query processing. *ACM Computing Surveys (CSUR)*, 50(5):64, 2017.

[4] R. Bača, M. Krátký, T. W. Ling, and J. Lu. Optimal and efficient generalized twig pattern processing: a combination of preorder and postorder filterings. *The VLDB Journal*, 22(3):369–393, 2013.

[5] R. Bača, P. Lukáš, and M. Krátký. Cost-based holistic twig joins. *Information Systems*, 52:21–33, 2015.

[6] M. Brantner, S. Helmer, C.-C. Kanne, and G. Moerkotte. Full-fledged algebraic XPath processing in Natix. In *Proceedings of Data Engineering, 2005. ICDE 2005.*, pages 705–716. IEEE, 2005.

[7] N. Bruno, N. Koudas, and D. Srivastava. Holistic twig joins: optimal xml pattern matching. In *Proceedings of the 2002 ACM SIGMOD international conference on Management of data*, pages 310–321. ACM, 2002.

[8] S. Chen, H.-G. Li, J. Tatemura, W.-P. Hsiung, D. Agrawal, and K. S. Candan. Twig2Stack: Bottom-up Processing of Generalized-tree-pattern Queries Over XML documents. In *Proceedings of VLDB 2006*, pages 283–294, 2006.

[9] T. Chen, J. Lu, and T. W. Ling. On Boosting Holism in XML Twig Pattern Matching Using Structural Indexing Techniques. In *Proceedings of ACM SIGMOD 2005*, pages 455–466. ACM Press, 2005.

[10] Z. Chen, H. V. Jagadish, L. V. S. Lakshmanan, and S. Paparizos. From tree patterns to generalized tree patterns: on efficient evaluation of XQuery. *Proceedings of the 29th international conference on Very Large Data Bases, VLDB 2003*, pages 237–248, 2003.

[11] Database Research Group. RadegastXDB, http://db.cs.vsb.cz/Projects, 2017.

[12] U. Dayal. Of nests and trees: A unified approach to processing queries that contain nested subqueries, aggregates, and quantifiers. In *Proceedings of the 13th International Conference on Very Large Data Bases*, pages 197–208. Morgan Kaufmann Publishers Inc., 1987.

[13] M. Fernández, J. Hidders, P. Michiels, J. Siméon, and R. Vercammen. Optimizing sorting and duplicate elimination in xquery path expressions. In *International Conference on Database and Expert Systems Applications (DEXA)*, pages 554–563. Springer, 2005.

[14] M. Fontoura, V. Josifovski, E. Shekita, and B. Yang. Optimizing cursor movement in holistic twig joins. In *Proceedings of the 14th ACM international conference on Information and knowledge management*, pages 784–791. ACM, 2005.

[15] H. Georgiadis, M. Charalambides, and V. Vassalos. Efficient physical operators for cost-based xpath execution. In *Proceedings of the 13th International Conference on Extending Database Technology*, pages 171–182. ACM, 2010.

[16] G. Gottlob, C. Koch, and R. Pichler. Efficient algorithms for processing xpath queries. *ACM Transactions on Database Systems (TODS)*, 30(2):444–491, 2005.

[17] N. Grimsmo, T. A. Bjørklund, and M. L. Hetland. Fast Optimal Twig Joins. In *Proceedings of the 36th International Conference on Very Large Data Bases, VLDB 2010*, pages 894–905. VLDB Endowment, 2010.

[18] J. Lu, T. Chen, and T. W. Ling. Efficient Processing of XML Twig Patterns with Parent Child Edges: a Look-ahead Approach. In *Proceedings of ACM CIKM 2004*, pages 533–542. ACM Press, 2004.

[19] J. Lu, T. W. Ling, Z. Bao, and C. Wang. Extended xml tree pattern matching: theories and algorithms. *IEEE transactions on knowledge and data engineering*, 23(3):402–416, 2011.

[20] J. Lu, T. W. Ling, C.-Y. Chan, and T. Chen. From Region Encoding to Extended Dewey: on Efficient Processing of XML Twig Pattern Matching. In *Proceedings of the 31st International Conference on Very Large Data Bases, VLDB 2005*, pages 193–204, 2005.

[21] P. Lukáš, R. Bača, and M. Krátký. Cooking lightweight xml query processor with binary joins and comparing it with holistic joins: Technical report. *arXiv*, abs/1703.09539, 2017.

[22] C. Luo, Z. Jiang, W.-C. Hou, F. Yu, and Q. Zhu. A sampling approach for xml query selectivity estimation. In *Proceedings of the 12th International Conference on Extending Database Technology: Advances in Database Technology*, pages 335–344. ACM, 2009.

[23] C. Mathis and T. Härder. Hash-based structural join algorithms. In *International Conference on Extending Database Technology*, pages 136–149. Springer, 2006.

[24] P. Michiels, G. Mihaila, and J. Siméon. Put a Tree Pattern in Your Algebra. In *Proceedings of the 23th International Conference on Data Engineering, ICDE 2007*, pages 246–255, 2007.

[25] S. Paparizos, Y. Wu, L. V. S. Lakshmanan, and H. V. Jagadish. Tree logical classes for efficient evaluation of XQuery. In *Proceedings of the 2004 ACM SIGMOD international conference on Management of data*, pages 71–82. ACM, 2004.

[26] N. Polyzotis and M. Garofalakis. Xsketch synopses for xml data graphs. *ACM Transactions on Database Systems (TODS)*, 31(3):1014–1063, 2006.

[27] L. Qin, J. X. Yu, and B. Ding. TwigList: Make Twig Pattern Matching Fast. In *The 12th International Conference on Database Systems for Advanced Applications, DASFAA 2007*, volume 4443 of *LNCS*, pages 850–862. Springer-Verlag, 2007.

[28] J. Robie, D. Chamberlin, M. Dyck, and J. Snelson. Xquery 3.0: An xml query language, w3c recommendation. *Retrieved September*, 7:2017, 2014.

[29] A. Schmidt, F. Waas, M. Kersten, M. J. Carey, I. Manolescu, and R. Busse. Xmark: A benchmark for xml data management. In *Proceedings of the 28th international conference on Very Large Data Bases*, pages 974–985. VLDB Endowment, 2002.

[30] S. Son, H. Shin, and Z. Xu. Structural semi-join: A light-weight structural join operator for efficient xml path query pattern matching. In *Database Engineering and Applications Symposium, 2007. IDEAS 2007. 11th International*, pages 233–240. IEEE, 2007.

[31] I. Tatarinov, S. D. Viglas, K. Beyer, J. Shanmugasundaram, E. Shekita, and C. Zhang. Storing and querying ordered xml using a relational database system. In *Proceedings of the 2002 ACM SIGMOD international conference on Management of data*, pages 204–215. ACM, 2002.

[32] University of Washington Database Group. The XML Data Repository, http://www.cs.washington.edu/research/xmldatasets/, 2002.

[33] A. M. Weiner and T. Härder. Using structural joins and holistic twig joins for native xml query optimization. In *East European Conference on Advances in Databases and Information Systems*, pages 149–163. Springer, 2009.

[34] A. M. Weiner and T. Härder. An integrative approach to query optimization in native XML database management systems. In *Proceedings of the Fourteenth International Database Engineering & Applications Symposium*, IDEAS '10, pages 64–74, New York, NY, USA, 2010. ACM.

[35] H. Wu, T. W. Ling, and G. Dobbie. Tp+ output: modeling complex output information in xml twig pattern query. In *International XML Database Symposium*, pages 128–143. Springer, 2010.

[36] Y. Wu, J. M. Patel, and H. Jagadish. Structural join order selection for xml query optimization. In *Data Engineering, 2003. Proceedings. 19th International Conference on*, pages 443–454. IEEE, 2003.

[37] M. Yoshikawa, T. Amagasa, T. Shimura, and S. Uemura. XRel: a Path-based Approach to Storage and Retrieval of XML Documents Using Relational Databases. *ACM Transactions on Internet Technology*, pages 110–141, 2001.

[38] C. Zhang, J. Naughton, D. DeWitt, Q. Luo, and G. Lohman. On supporting containment queries in relational database management systems. In *ACM SIGMOD Record*, volume 30, pages 425–436. ACM, 2001.

# APPENDIX A
# PROBLEMATIC ORDER OF FOR CLAUSES

In Figure 16a, we see a query Q8 for which there is not any TPQ whose pre-order traversal of the output query nodes matches the order of the 'for' clauses. If we processed Q8 by a TPQ in Figure 16b, the result would be sorted by (#a, #b, #d, #c) and an additional sorting would be necessary. To the best of our knowledge, in the literature, there is no binary or holistic join approach with focus on the processing of such queries without an additional sorting of the result.

```
for $a in //a
for $b in $a/b
for $c in $a/c
for $d in $b/d
return  ($a, $b, $c, $d)
```

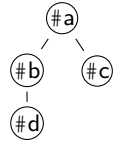

(a) Query Q8        (b) TPQ of Q8

Fig. 16. An XQuery query with a problematic order of 'for' clauses

# APPENDIX B
# PROPERTIES OF A QUERY CORE

In this section, we prove that in a query core, each non-output query node has exactly one child. This follows from a property of a real XQuery query: for each pair of output query nodes, their lowest common ancestor is also an output query node.

We give a proof by contradiction. Let $q_i$ be a non-output query core node. Let $q_a$ and $q_b$ be query core nodes such that $q_i$ is their parent. Then $q_a$ and $q_b$ lie on a path between two output query nodes (see the definition of a query core in Section 2). Therefore, in subtrees of $q_a$ and $q_b$ there must be output query nodes $q_{o_a}$ and $q_{o_b}$, respectively. However, $q_i$ is the lowest common ancestor of $q_{o_a}$ and $q_{o_b}$ and, therefore, it has to be an output query node.

# APPENDIX C
# SUBOPTIMAL QUERY PROCESSING FOR RECURSIVE STREAM

In this section, we discuss an example where the binary join approach is not optimal to process a query because of a recursive stream $T_a$ and the unbounded size of the self- and inherited-lists of the *StackTreeAnc* algorithm. Let us process the query Q9 in Figure 17b on the document in Figure 17a. The plan of Q9 is depicted in Figure 17c and we see that it consists only of one partial-join algorithm *StackTreeAnc*.

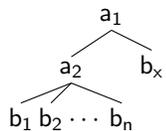      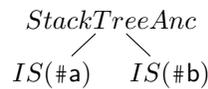

(a) XML document    (b) Query Q9    (c) Plan of Q9

Fig. 17. Suboptimal query processing

The operator starts with reading the data nodes $a_1$ and $a_2$. Since $a_2$ is a descendant of $a_1$, the stream $T_a$ is recursive and both data nodes are put on the stack. After that, the algorithm reads the data nodes $b_1 \cdots b_n$ and it determines that all pairs $[a_2, b_1] \cdots [a_2, b_n]$ should be output. However, these pairs cannot be output until $a_1$ is popped out from the stack. This is because of the possibility that there is a data node $b_x$ following $a_2$ that will be joined with $a_1$. Since *StackTreeAnc* is an ancestor-sorted partial-join, it has to produce the result sorted by ancestors, i.e., it has to first output all pairs joined with $a_1$. Therefore, $[a_2, b_1] \cdots [a_2, b_n]$ are not output immediately. Instead, they are placed into the self-list of $a_2$. The situation is depicted in Figure 18.

| $a_2$ | self-list: $\{[a_2, b_1] \cdots [a_2, b_n]\}$, inherited-list: $\varnothing$ |
| $a_1$ | self-list: $\varnothing$, inherited-list: $\varnothing$ |
| stack | |

Fig. 18. StackTreeAnc after reading of $b_n$

After the reading of $b_n$, $a_2$ can be popped out from the stack and its self-list is added into the inherited-list of $a_1$. Subsequently, $b_x$ is read and $[a_1, b_x]$ is then added into the self-list of $a_1$. Finally, $a_1$ is popped out from the stack and its self-list and inherited-list are enumerated to the output, respectively.

In Figure 18, we can see that there are possibly many pairs $[a_2, b_1] \cdots [a_2, b_n]$ that have to be stored in the self-list of $a_2$. The number of pairs is not bounded by the depth of the document, therefore, the binary join approach is not optimal to process the query Q9 on the document in Figure 17a.

# APPENDIX D
# PROCESSING OF GTP

In [10] a *generalized twig pattern* (GTP) is introduced. Such a query model supports the XQuery semantics in many aspects. Apart from a specification of output and non-output query nodes, it defines aggregation query nodes which correspond to 'let' clauses. Moreover, it supports boolean formulas with 'or' and 'not' connectives. In this section, we discuss how to extend our binary join approach to support such a model. Let us note that RXDB (our prototype of a native XML database) supports the GTP.

## D.1 'let' Clauses

Let us call *aggregation query nodes* the query nodes which correspond to the 'let' clauses. In Figure 19a, we see an example of such a query. The corresponding GTP is depicted in Figure 19b, where the aggregation query nodes are squared, i.e., #e is the only aggregation query node.

The processing of the aggregation query nodes is similar to the processing of output query nodes. However, in a principle, there are two differences:

1) In Section 2.5, we defined a table as an array of tuples of data nodes. When the aggregation query nodes are considered, the columns corresponding to such query nodes contain sets of data nodes. The result of Q10 on the document in Figure 1a is depicted in Table 9. We see, the cells in the column #e contain sets of data nodes.

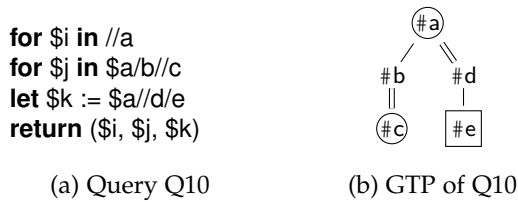

(a) Query Q10        (b) GTP of Q10

Fig. 19. An XQuery query with a 'let' clause

| #a | #c | #e |
|----|----|----|
| $a_1$ | $c_1$ | $e_1, e_2$ |
| $a_1$ | $c_2$ | $e_1, e_2$ |
| $a_2$ | $c_3$ | $e_3$ |

TABLE 9
The result of the query Q10

2) Since the sets of data nodes corresponding to the aggregation query nodes can be empty, we need to introduce an outer partial-join. Otherwise we could loose some of the results of the 'for' clauses. In Line 20 of *BuildPlanCore* (see Algorithm 3), if $c$ is an aggregation query node, we need to introduce an outer version of the ancestor-sorted partial-join *StackTreeAncSrt*. The implementation the partial-join is straightforward: all tuples of $T_a$ are always retained, even if there are no corresponding tuples in $T_d$.

### D.2  Boolean Formulas

The TPQ model defined in Section 2.1 can be used to process only XQuery queries with conjunctive boolean formulas, i.e., boolean formulas where only 'and' connectives appear. However, in many real-world queries, we can also find 'or' and 'not' connectives. An example of such a query can be found in Figure 20a. In the corresponding GTP (see Figure 20b), we see that #d is labeled with a boolean formula that has to be satisfied between #f and #c.

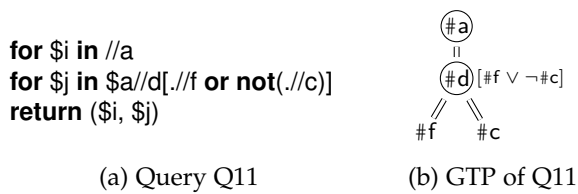

(a) Query Q11        (b) GTP of Q11

Fig. 20. An XQuery query with a 'let' clause

To process a query node $q$ labeled with a boolean formula, we introduce a generalized version of *FilterAncPC* such that it can have more input streams $T_d$. Each $T_d$ corresponds to one child query node of $q$, nevertheless, it still has one input stream $T_a$. Before a data node $a$ of $T_a$ is output, we check whether the boolean formula is satisfied between the corresponding data nodes of the streams $T_d$. We can observe that such an approach brings some similarities with a holistic join algorithm, since the generalized version of *FilterAncPC* processes more than two streams at once.

1919# APPENDIX E
# TESTING TWIG PATTERNS

For each twig pattern $p$ in this appendix, we randomly generated $n$ queries, where $n$ is the number of query nodes in $p$, such that each query has a different number of output query nodes.

## E.1 XMark

| | |
|---|---|
| XM01 | //categories/category[//description//text and /@id] |
| XM02 | //closed_auctions/closed_auction[//price]/buyer/@person |
| XM03 | //regions/samerica/item[/mailbox/mail/to]/incategory/@category |
| XM04 | //open_auctions//open_auction[/reserve and //bidder/personref/@person]/itemref/@item |
| XM05 | //open_auctions/open_auction[/interval//start] |
| XM06 | //regions/namerica/item[//from]/incategory/@category |
| XM07 | //regions[//europe]/namerica/item/description[/parlist]//listitem/text |
| XM08 | //people/person[/emailaddress and //creditcard]/@id |
| XM09 | //closed_auctions/closed_auction[/type]/quantity |
| XM10 | //people/person/profile[//business]/interest/@category |
| XM11 | //closed_auctions//annotation[//happiness and /description/text] |
| XM12 | //closed_auctions//closed_auction[//date]//itemref[/@item] |
| XM13 | //closed_auctions/closed_auction[//buyer]/itemref[/@item] |
| XM14 | //closed_auctions/closed_auction[//seller/@person]/annotation |
| XM15 | //open_auctions/open_auction[/annotation/description//text]//date |
| XM16 | //open_auctions/open_auction//bidder[/increase]/date |
| XM17 | //open_auctions/open_auction[/@id]/bidder/date |
| XM18 | //regions//item[//text/keyword]/mailbox/mail[/to]/from |
| XM19 | //regions[/samerica]//item[/description/text]/incategory[/@category] |
| XM20 | //regions//item[//bold and /payment]//shipping |
| XM21 | //regions//item[/description/text and /location]/shipping |
| XM22 | //open_auctions//open_auction[/current and //bidder]/initial |
| XM23 | //open_auctions/open_auction[/bidder]//interval |
| XM24 | //open_auctions/open_auction[/bidder]//initial |
| XM25 | //open_auctions//open_auction[//bidder[/time]/date]/initial |
| XM26 | //open_auctions/open_auction[/bidder[/personref/@person]/time]/interval//end |
| XM27 | //categories/category[/description/text//emph and /@id] |
| XM28 | //categories/category[//name and /@id]/description//text |
| XM29 | //categories/category[/name]/description/parlist/listitem/text |
| XM30 | //people/person[/@id and /homepage and //name] |
| XM31 | //open_auctions/open_auction[/reserve]//seller/@person |
| XM32 | //categories/category[/description/text/keyword]/name |
| XM33 | //regions//samerica/item[/incategory/@category and //name]/shipping |
| XM34 | //open_auctions/open_auction[/interval/start]/seller/@person |
| XM35 | //categories/category[//description/parlist/listitem] |
| XM36 | //open_auctions/open_auction[/quantity and /current]/annotation[/description/text[/bold]/emph]/happiness |
| XM37 | //categories/category/description/parlist[//listitem/text/emph] |
| XM38 | //categories/category[//name]/description[/text] |
| XM39 | //closed_auctions/closed_auction[/seller/@person]/annotation[/happiness]//description[/parlist/listitem/text//emph/bold] |
| XM40 | //categories/category/description//parlist/listitem//text/bold |
| XM41 | //people//person[/homepage and /@id]/profile/business |
| XM42 | //people/person[/@id and /homepage and /creditcard] |
| XM43 | //regions//australia/item[//name and //description//keyword]/incategory/@category |
| XM44 | //regions[/europe]/africa/item[/mailbox/mail[//text]/from and /payment]/incategory/@category |
| XM45 | //categories/category[/@id and //parlist/listitem/text/bold] |
| XM46 | //categories/category[//name and /description/text//emph and /@id] |
| XM47 | //open_auctions/open_auction[/bidder[//time and //increase]//date and //reserve] |
| XM48 | //closed_auctions/closed_auction[/type and /quantity]/annotation/author/@person |
| XM49 | //categories/category[/name]/description/text[/emph] |
| XM50 | //categories/category[/@id and //description/parlist/listitem//text/bold] |
| XM51 | //item[//bold and //emph/bold] |
| XM52 | //item[//emph and //keyword//emph] |
| XM53 | //item[//keyword//bold and //bold] |
| XM54 | //item[//bold/keyword and //keyword] |
| XM55 | //item[/incategory and //bold//emph] |
| XM56 | //item[//keyword and //emph//keyword] |
| XM57 | //item[//emph//bold and //keyword] |
| XM58 | //item[//keyword//emph and //bold] |
| XM59 | //item[//emph and //keyword/bold] |
| XM60 | //item[//emph and //bold/emph] |

## E.2  SwissProt

```
SP01    //Entry[//MIM/@sec_id and /@class and //Keyword]
SP02    //Entry[/@id]/HSSP[/@prim_id]/@sec_id
SP03    //Entry[/Org]/Features[/TRANSMEM[/@from]/@to]
SP04    //Entry[/Features//CHAIN[/@from]/Descr]/Keyword
SP05    //Entry[/Org and /Descr]//EMBL[/@sec_id]/@prim_id
SP06    //Entry[/Org and //HSSP/@prim_id]/@class
SP07    //Entry[/@class and //PIR/@prim_id]//Keyword
SP08    //Entry[/PIR[/@prim_id]/@sec_id and //Org]
SP09    //Entry[/PIR/@sec_id]/EMBL[/@prim_id]
SP10    //Entry[/PROSITE[/@status]/@sec_id]/Org
SP11    //Entry[//PROSITE/@status and /Keyword]/EMBL[/@prim_id]
SP12    //Entry[/@class]/PIR[/@sec_id]/@prim_id
SP13    //Entry[/@mtype]/PRINTS[/@sec_id]/@prim_id
SP14    //Entry[/PIR[/@prim_id]/@sec_id and /Keyword]//Features[/TRANSMEM[//Descr]/@to]
SP15    //Entry[/PDB/@sec_id]/Features[//Descr and /CONFLICT/@from]
SP16    //Entry[/@id]/Features[/DOMAIN[/Descr]/@from]//VARSPLIC/@to
SP17    //Entry/Features[/NP_BIND]//CHAIN[/Descr]/@from
SP18    //Entry[/@mtype and /Features/BINDING/@to]
SP19    //Entry[/Org and /PFAM[/@status]/@prim_id]//Features/TRANSIT/Descr
SP20    //Entry[//REPEAT]/INTERPRO[/@sec_id and /@prim_id]
SP21    //Entry[/Mod/@Rel]/Features/CHAIN[/Descr and /@from]/@to
SP22    //Entry[//Ref]//PIR[/@prim_id]/@sec_id
SP23    //Entry[/PFAM/@status and /INTERPRO/@sec_id]//Features[//MOD_RES/@to]
SP24    //Entry[/Descr]/Features[/SITE/@from]
SP25    //Entry[/TIGR/@prim_id and /EMBL/@sec_id]
SP26    //Entry[/Species and /SWISS-2DPAGE/@prim_id]
SP27    //Entry[/INTERPRO]/Features/ACT_SITE[/@from]/Descr
SP28    //Entry[/Species]/TIGR[/@sec_id]/@prim_id
SP29    //Entry[/Org]//Features[/NON_TER/@to]/METAL//Descr
SP30    //Entry[/Features[/DISULFID[/@to]/Descr]/CHAIN]/HSSP/@prim_id
SP31    //Entry[/Features[/CHAIN]/TRANSIT/@from]
SP32    //Entry[/@mtype and //SUBTILIST[/@sec_id]/@prim_id]
SP33    //Entry[/Mod[/@type]/@date]/PRINTS/@prim_id
SP34    //Entry[/FLYBASE[/@prim_id]/@sec_id and /Gene]/@seqlen
SP35    //Entry[/Org]/Features//NP_BIND/@from
SP36    //Entry[/Features//METAL[/@to]/@from and /Species]/@class
SP37    //Entry[/@class and /EMBL]/HSSP/@sec_id
SP38    //Entry[/Org]//MGD[/@prim_id]/@sec_id
SP39    //Entry[/Org and /PDB/@sec_id]/@mtype
SP40    //Entry[/PFAM/@status and /MGD/@sec_id]
SP41    //Entry[//Keyword]//PRINTS[/@sec_id]/@prim_id
SP42    //Entry[//TIGR[/@prim_id]/@sec_id]/Org
SP43    //Entry[//Keyword and /PROSITE[/@sec_id and /@status]/@prim_id]//Features//ACT_SITE[/@to]/@from
SP44    //Entry[//INTERPRO/@sec_id]/PIR/@prim_id
SP45    //Entry[/Org]/HSSP[/@sec_id]/@prim_id
SP46    //Entry[//TUBERCULIST]/Mod[/@date and /@Rel]
SP47    //Entry[/Mod[/@type and /@Rel]/@date]/PIR/@prim_id
SP48    //Entry[//Org]//Features/METAL[/@from]/Descr
SP49    //Entry[/Keyword]/GCRDB[/@sec_id]/@prim_id
SP50    //Entry[/Mod/@date]//PRINTS[/@prim_id]/@sec_id
SP51    //Entry[//Ref/MedlineID and //Features]
SP52    //Entry[/Ref//Comment and //Descr]
SP53    //Entry[//Ref/Comment and //Descr]
SP54    //Entry//Ref[/Author and //Cite]
SP55    //Entry[//Features//Descr and //PROSITE]
SP56    //Entry//Ref[/Author and /MedlineID]
SP57    //Entry[/Ref//Author and /PFAM]
SP58    //Entry[//Descr and //Ref[/Cite and /Comment]]
SP59    //Entry[/Organelle and //CA_BIND//Descr]
SP60    //Entry[//SIGNAL/Descr and /FLYBASE]
```

### E.3 TreeBank

```
TB01   //EMPTY[//S[/_BACKQUOTES_]//VP/VBZ]//_PERIOD_
TB02   //EMPTY[//VBN and //_PERIOD_]//NP[//DT]//NN
TB03   //EMPTY[/S[//VBD and //JJ]/VP/VBN and //PP/IN]
TB04   //EMPTY[/_PERIOD_ and /S[/VBP]//TO]
TB05   //EMPTY[/S[//VP//NP[/DT]/RB and //_NONE_]
TB06   //EMPTY[/S and //PP[/IN]/NP[/VBN]/NN]
TB07   //EMPTY[/S/VP/SBAR//_NONE_]/_PERIOD_
TB08   //EMPTY[//_PERIOD_ and /S/PP[/IN]/NP/NNS]
TB09   //EMPTY[//_PERIOD_]/S[//VBD]/NP[/DT]
TB10   //EMPTY[//_PERIOD_]//S[/NP/ADJP//JJ]/VB
TB11   //EMPTY[/S[/NP[/DT]/NNS and /RB]/VP]
TB12   //EMPTY/S[//NP/NNS and /VP and //RB]
TB13   //EMPTY[//_PERIOD_]/S[/NP[//CC]//NNS]//VBP
TB14   //EMPTY/S[/_PERIOD_ and /NP//NN]//ADJP/PP[//IN and //PRP_DOLLAR_]
TB15   //EMPTY//S[/NP/PRP and //_PERIOD_]
TB16   //EMPTY[/_PERIOD_]//S/VP[//VBD]//PP[//TO]//NP[//_NONE_]/CD
TB17   //EMPTY[//_PERIOD_]/_QUESTIONMARK_[/S//_NONE_]
TB18   //EMPTY[/S[//NNP]/VP[//VBD]/NP//CD]
TB19   //EMPTY//SINV[/_COMMA_ and //_PERIOD_]
TB20   //EMPTY//S//SBAR[/CC and //_NONE_]
TB21   //EMPTY[/_PERIOD_]/S[/CC and /_COMMA_]
TB22   //EMPTY[//S[//_PERIOD_ and //ADJP/PP//IN]/VP//VB and //MD]
TB23   //EMPTY[/S[/ADJP[//JJ]//RB]/VP/VBZ]
TB24   //EMPTY[//_PERIOD_]/S[/VBP]//VP//VBN
TB25   //EMPTY/S[/VP/PP//TO and /_PERIOD_]
TB26   //EMPTY//S[//VBD and /VP/PP/IN]/NP[/NN]
TB27   //EMPTY[//_PERIOD_]/S[/VBZ]//PP/NP[//JJ]/NN
TB28   //EMPTY[//_PERIOD_]/S[//SBAR/_NONE_]/VP
TB29   //EMPTY//S[//VB and //RB]/NP[//JJ and /PP/IN]//_QUOTES_
TB30   //EMPTY[/_PERIOD_]/S[/NP[/NNS]//NNP]//ADJP[/RB]
TB31   //EMPTY/S[/_PERIOD_]/NP[//NNP and //_LRB_]
TB32   //EMPTY[/VP//ADJP/JJ]//S//_PERIOD_
TB33   //EMPTY[/_PERIOD_]/SBAR[//IN]//S[//NNP and //VP[/VBZ]//PP//NN]
TB34   //EMPTY[/_PERIOD_]//S//VP[/VBZ]/PP[/IN]/NP//ADJP/JJ
TB35   //EMPTY[/_PERIOD_]/S[//VP[/VBN]//TO]/NP[//PP/IN]/NNPS
TB36   //EMPTY[/S//VP[//VB]/PP[//VBG]/TO]
TB37   //EMPTY[/_PERIOD_]/S[/VP[/VBZ]/NP and //_COMMA_]
TB38   //EMPTY[//_PERIOD_]/S[//VP[/NP/NN]/VBD and //SBAR]//VP-1/VBG
TB39   //EMPTY[/_PERIOD_]/S/NP[/CD and /NNS]
TB40   //EMPTY[//_PERIOD_ and /S[/NP[//JJ]/NN]/RB]
TB41   //EMPTY[/S[//VP//VB and /VBZ]/RB and /_PERIOD_]
TB42   //EMPTY[/_PERIOD_]//S/SINV/VP[//VBZ]
TB43   //EMPTY[/_PERIOD_]/S[/NP/PRP]/PP//IN
TB44   //EMPTY[/S[/VP[/PP[/NP/NN]//IN]/RB]//_PERIOD_]
TB45   //EMPTY[//_PERIOD_]//S/SBARQ[/SQ]/_NONE_
TB46   //EMPTY//S[//_BACKQUOTES_]/NP[/PRP]
TB47   //EMPTY[//X[//_COLON_]//NP//NNPS]/_PERIOD_
TB48   //EMPTY[/_PERIOD_ and /S[/PP[/NP/NNP]//CC]/_COMMA_]
TB49   //EMPTY[/_PERIOD_]/S/SINV/VP[//VBD]
TB50   //EMPTY[/_PERIOD_]//S[/VP[//VBD]//PP/IN]/NP[//NNS]/DT
TB51   //EMPTY//PP/NP//NNS
TB52   //EMPTY[//_NONE_ and //VP/NP]
TB53   //EMPTY//VP/NP//NNS
TB54   //EMPTY//NP[/JJ and //NN]
TB55   //EMPTY[//NP//_NONE_ and //IN]
TB56   //EMPTY[//NP and //PP//NN]
TB57   //EMPTY[//VP//DT and //NP]
TB58   //EMPTY[//NN and //PP/IN]
TB59   //EMPTY[//NN and /S//NNS]
TB60   //EMPTY[//NNS and //VP//NN]
```

# Appendix F
# Queries for Performance Comparison

## F.1 XMark

1.  ```
    for $a in //categories/category[.//description[.//text] and ./@id]
    return ($a)
    ```

2.  ```
    for $a in //closed_auctions/closed_auction[.//price and ./buyer[./@person]]
    return ($a)
    ```

3.  ```
    for $a in //regions/samerica/item[./mailbox[./mail[./to]]]/incategory/@category
    return ($a)
    ```

4.  ```
    for $a in //open_auctions//open_auction[./reserve and .//bidder[./personref[./@person]]]/itemref[./@item]
    return ($a)
    ```

5.  ```
    for $a in //regions/namerica/item[.//from]/incategory/@category
    return ($a)
    ```

6.  ```
    for $a in //regions
    for $b in $a/namerica
    for $c in $b/item
    for $d in $c//from
    for $e in $c/incategory
    for $f in $e/@category
    return ($a, $b, $c, $d, $e, $f)
    ```

7.  ```
    for $a in //regions
    for $b in $a//europe
    for $c in $a/namerica
    for $d in $c/item
    for $e in $d/description
    for $f in $e/parlist
    for $g in $e//listitem
    for $h in $g/text
    return ($a, $b, $c, $d, $e, $f, $g, $h)
    ```

8.  ```
    for $a in //people
    for $b in $a/person
    for $c in $b/emailaddress
    for $d in $b//creditcard
    for $e in $b/@id
    return ($a, $b, $c, $d, $e)
    ```

9.  ```
    for $a in //closed_auctions
    for $b in $a/closed_auction
    for $c in $b/type
    for $d in $b/quantity
    return ($a, $b, $c, $d)
    ```

10. ```
    for $a in //people
    for $b in $a/person
    for $c in $b/profile
    for $d in $c//business
    for $e in $c/interest
    for $f in $e/@category
    return ($a, $b, $c, $d, $e, $f)
    ```

### F.2 TreeBank

1. for $a in //EMPTY[.//S[./_BACKQUOTES_ and .//VP[./VBZ]]]//_PERIOD_
   return ($a)

2. for $a in //EMPTY[.//PP[./IN]]/S[.//VBD and .//JJ]/VP[./VBN]
   return ($a)

3. for $a in //EMPTY[./_PERIOD_ and ./S[./VBP and .//TO]]
   return ($a)

4. for $a in //EMPTY[./S]//PP[./IN]/NP[./VBN and ./NN]
   return ($a)

5. for $a in //EMPTY[.//_PERIOD_]/S[./PP[./IN and ./NP[./NNS]]]
   return ($a)

6. for $a in //EMPTY
   for $b in $a/S
   for $c in $b//NP
   for $d in $c/NNS
   for $e in $b/VP
   for $f in $b//RB
   return ($a, $b, $c, $d, $e, $f)

7. for $a in //EMPTY
   for $b in $a//_PERIOD_
   for $c in $a//S
   for $d in $c/NP
   for $e in $d//CC
   for $f in $d//NNS
   for $g in $c//VBP
   return ($a, $b, $c, $d, $e, $f, $g)

8. for $a in //EMPTY
   for $b in $a//S
   for $c in $b/NP
   for $d in $c/PRP
   for $e in $b//_PERIOD_
   return ($a, $b, $c, $d, $e)

9. for $a in //EMPTY
   for $b in $a/_PERIOD_
   for $c in $a//S
   for $d in $c//VP
   for $e in $d//VBD
   for $f in $d//PP
   for $g in $f//TO
   for $h in $f//NP
   for $i in $h//_NONE_
   for $j in $h/CD
   return ($a, $b, $c, $d, $e, $f, $g, $h, $i, $j)

10. for $a in //EMPTY
    for $b in $a/S
    for $c in $b//NNP
    for $d in $b/VP
    for $e in $d//VBD
    for $f in $d/NP
    for $g in $f//CD
    return ($a, $b, $c, $d, $e, $f, $g)